\documentclass[final,3p,times]{elsarticle}

\usepackage{amssymb}
\usepackage{subfigure}
\usepackage{algorithm}
\usepackage{algpseudocode}

\usepackage{multirow}
\usepackage{amsmath}
\usepackage{xcolor}
\usepackage{color}
\definecolor{webgreen}{rgb}{0,.5,0}
\usepackage{colortbl}
\usepackage{graphicx}
\usepackage{booktabs}
\usepackage{url}
\usepackage{subfigure}
\usepackage[titletoc]{appendix}
\usepackage{diagbox}
\usepackage{caption}
\usepackage{natbib}
\usepackage{dsfont}

\newcommand{\rmnum}[1]{\romannumeral #1}
\newcommand{\Rmnum}[1]{\expandafter\@slowromancap\romannumeral #1@}

\definecolor{pblue}{rgb}{0.13,0.13,1}
\definecolor{pgreen}{rgb}{0,0.5,0}
\definecolor{pred}{rgb}{0.9,0,0}
\definecolor{pgrey}{rgb}{0.46,0.45,0.48}
\definecolor{ppurple}{rgb}{102, 0, 204}
\usepackage{listings}
\usepackage{wrapfig}
\biboptions{authoryear}
\usepackage[colorlinks,linkcolor=blue,citecolor=blue,urlcolor=blue]{hyperref}
\usepackage{fancyhdr}
\def\WithComments{}
\ifdefined \WithComments

\else

\fi
\usepackage{soul}

\journal{Computers \& Security }

\begin{document}
\begin{frontmatter}
\title{Automated Poisoning Attacks and Defenses in Malware Detection Systems:  \\An Adversarial Machine Learning Approach}
\author[label1,label2]{Sen Chen}
\address[label1]{East China Normal University, Shanghai, China}
\address[label2]{Nanyang Technological University, Singapore\fnref{label4}}

\cortext[cor1]{Corresponding author. Email address: lhxu@cs.ecnu.edu.cn. \\We would like to thank Pwnzen Infotech Inc. for providing us with a copy of mobile malware to conduct the study, especially the Pwnzen Infotech Inc. co-founder Zhushou Tang for exchanging helpful industry experience. This work was supported in part by the National Natural Science Foundation of China, under Grant 61502170, 61272444, 61411146001, U1401253, and U1405251, in part by the Science and Technology Commission of Shanghai Municipality under Grant 13ZR1413000. }
\ead{ecnuchensen@gmail.com}

\author[label3,label4]{Minhui Xue}
\address[label3]{ New York University Shanghai, Shanghai, China}
\address[label4]{Shanghai Jiao Tong University, Shanghai, China}
\ead{minhuixue@nyu.edu}

\author[label1,label2]{Lingling Fan}
\ead{ecnujanefan@gmail.com}

\author[label5]{Shuang Hao}
\ead{shao@utdallas.edu}
\address[label5]{University of Texas at Dallas, USA}

\author[label1]{Lihua Xu\corref{cor1}}
\ead{lhxu@cs.ecnu.edu.cn}

\author[label4]{Haojin Zhu}
\ead{zhuhaojin@gmail.com}

\author[label6]{Bo Li}
\address[label6]{University of California, Berkeley, USA}
\ead{lxbosky@gmail.com }

\begin{abstract}
The evolution of mobile malware poses a serious threat to smartphone security. Today, sophisticated attackers can adapt by maximally sabotaging machine-learning classifiers via polluting training data, rendering most recent machine learning-based malware detection tools (such as \textsc{Drebin}, \textsc{DroidAPIMiner}, and \textsc{MaMaDroid}) ineffective. In this paper, we explore the feasibility of constructing crafted malware samples; examine how machine-learning classifiers can be misled under three different threat models; then conclude that injecting carefully crafted data into training data can significantly reduce detection accuracy. To tackle the problem, we propose \textsc{KuafuDet}, a two-phase learning enhancing approach that learns mobile malware by adversarial detection. \textsc{KuafuDet} includes an offline training phase that selects and extracts features from the training set, and an online detection phase that utilizes the classifier trained by the first phase. To further address the adversarial environment, these two phases are intertwined through a self-adaptive learning scheme, wherein an automated camouflage detector is introduced to filter the suspicious false negatives and feed them back into the training phase. We finally show that \textsc{KuafuDet} can significantly reduce false negatives and boost the detection accuracy by at least 15\%. Experiments on more than 250,000 mobile applications demonstrate that \textsc{KuafuDet} is scalable and can be highly effective as a standalone system.
\end{abstract}

\end{frontmatter}

\section{Introduction}\label{sec:introduction}
Since last decade, the software development has been witnessed a massive shift toward mobile applications. With the growth of mobile applications and their users, the security and privacy concerns are increasingly becoming the focus of great concern to various stakeholders. For instance, more and more users store personal data in their mobile devices, even carrying out financial transactions such as online banking and shopping from their smartphones. Some of these data can be very sensitive. Consequently, hackers can have substantial financial gain from such sensitive data and thus find mobile devices to be lucrative targets.

It is not surprising that the demand for tools of automatically analyzing and detecting malicious applications has also grown. Most of the researchers' and practitioners' efforts in this area target the Android platform, the largest share of the mobile market. There has been a plethora of research in malware detection for Android. Static and dynamic analyses are two generic techniques primarily implemented by two approaches: signature-based~\citep{schlegel2011soundcomber, zhou2012hey, zhou2013fast} and behavior-based~\citep{yan2012droidscope, wu2014airbag, tam2015copperdroid, graziano2015needles, rasthofer2016harvesting}. Information flow analysis-based approach~\citep{arzt2014flowdroid, li2015iccta, gordon2015information, enck2014taintdroid, wong2016intellidroid} is also proposed to detect Android malware. We note that machine learning is one of the most promising techniques in detecting mobile malware~\citep{aafer2013droidapiminer, arp2014drebin, yang2014droidminer, zhang2014semantics, rasthofer2014machine, avdiienko2015mining, yang2015appcontext, dash2016droidscribe, chen2016stormdroid, meng2016mystique,fan2016poster,idrees2017pindroid,feizollah2017androdialysis}. However, machine learning approaches also have a weakness: they are susceptible to adversarial countermeasures by attackers aware of their use. First, through reverse-engineering, attackers may become aware of classifiers and their parameters used to evade detection. Second, more sophisticated attackers can actively tamper with the classifiers by injecting well-crafted data into training data. Therefore, with Android's policy of open-source kernel, malware writers can gain an in-depth understanding of the mobile platform, hence intentionally alter the training set to reduce or eliminate its detection efficacy.

To our knowledge, most up-to-date work using machine learning mainly focused on detection accuracy and assumed that feature extraction is considered in an ideal environment~\citep{chen2016more, mariconti2016mamadroid}. No evasion techniques were conducted in the feature space when using machine learning-based detection approach. In this paper, we consider a threat model within a specific class of attacks, named \textit{poisoning attack}, in which the attacker is assumed to control a subset of samples or inject additional seeds at will in order to mislead the learning algorithm. For example, in malware detection, a sophisticated attacker may have a good command of the whole training set and deliberately inject poisoning patterns to compromise the performance of the classifiers, which becomes more prone to misclassify malicious applications as benign ones. \emph{This threat model conceptually underlies adversarial machine learning: it involves gradients of the function $f$ represented by the learned model (\textit{e.g.}, SVM, logistic regression, $K$-Nearest Neighbor) in order to evaluate it on an input $x$. Attackers can then fully automatically either identify individual input features that are perturbed to achieve misclassification}~\citep{papernot2016limitations}. 
 
To test the ramifications of these causative attacks, we develop an adversarial model with three types of attackers according to different aggressiveness of attacks to simulate the real-world attacks. To do this, we adopt an customized adversarial crafting algorithm, characterized by the aggressiveness of attacks, to generate the crafted camouflage samples. The abstraction of crafting steps is somehow restricted in three ways. (\rmnum{1}) To preserve the functionality of the modified application, we only add or remove features; (\rmnum{2}) we add a restricted number of features. For simplicity, we therefore decide to add entries to the \textit{AndroidManifest.xml} and \textit{Smali} files; (\rmnum{3}) since obscuring semantic features is much more challenging than confusing syntax features, we only use syntax features to craft samples. In spite of these restrictions in crafting, we achieve a significantly high misclassification rate on malicious applications when using 564 original non-robust features. To further perform a longitudinal comparison, we also apply our poisoning attack to \textsc{Drebin}~\citep{arp2014drebin}, \textsc{DroidAPIMiner}~\citep{aafer2013droidapiminer}, and \textsc{MaMaDroid}~\citep{mariconti2016mamadroid}, the three most recent machine-learning detection systems in academia. We thus validate that the threat model and the poisoning attack are indeed viable in malware detection. We conjecture that almost all the state-of-the-art machine-learning malware detection systems are suffering from the poisoning attack we exhibited in the paper.

To handle these adversarial attacks, we propose \textsc{KuafuDet}, a learning enhancing defense system with adversarial detection that includes an offline training phase that selects and extracts contributing features from the training set for preprocessing, and an online detection phase that utilizes the classifier trained by the first phase. Comparing to existing work, these two phases act together, through a self-adaptive learning scheme, as an iterative adversarial detection process. Additionally, we introduce the \textit{camouflage detection} for verifying false alarms to protect against poisoning attacks. By using similarity analysis, the camouflage detection is applied to iteratively detect against malicious data distortion. In concrete, we train 16,000 Android application samples that are equally distributed, which are downloaded from Contagio Mobile Website,\footnote{\url{http://contagiominidump.blogspot.hk/}} Pwnzen Infotech Inc. and \textsc{Drebin}~\citep{arp2014drebin}. All 195 robust features are extracted using static analysis on the given application, pruned by information gain. We further evaluate the results on 4,000 applications. Our best detection classifier achieves up to 96\% accuracy without adversarial environment, and by at least 15\% accuracy when coping with the most powerful attackers, along with both low false negatives. Furthermore, we conduct an empirical evaluation on our test set and select 1,000 malware as samples out of the set of 10,400 malicious samples and scan them using \textsc{KuafuDet} and other industrial malware detecting tools, such as Kaspersky and McAfee encapsulated in VirusTotal.\footnote{\url{https://www.virustotal.com/}} The coverage of \textsc{KuafuDet} significantly outperforms these top-of-the-line antivirus systems. Finally, we perform the entire process of \textsc{KuafuDet}, using real-time streaming, on a server with 16 GB memory, quad-core i7-4800MQ at 3.6 GHz, and 1 TB hard drives and show that \textsc{KuafuDet} is scalable and efficient.
  
In this paper, we make the following key contributions that are fourfold. 
\begin{enumerate}
\item  We propose that poisoning attacks can be exhibited by three types of attackers in the real world, ranging from weak, strong, and sophisticated degrees. We hold evidence that the real-world mobile malware dataset is able to truly reflect three types of attackers we defined. 

\item We adopt a customized adversarial crafting algorithm, semantically characterized by the aggressiveness of attacks,  to generate the crafted camouflage samples using syntax features to largely simulate the real-world attacks. 

\item  We show that our poisoning attack is able to mislead \textsc{DroidAPIMiner}~\citep{aafer2013droidapiminer}, \textsc{Drebin}~\citep{arp2014drebin}, and \textsc{MaMaDroid}~\citep{mariconti2016mamadroid}, the three most recent machine-learning detection systems in academia.

\item We propose a two-phase iterative adversarial based detection, termed \textsc{KuafuDet}, wherein similarity-based filtering is used to identify the false negatives that are the camouflaged malicious applications, further reinforcing the resilience of the malware detection system.
\end{enumerate}

Our experiments show that attackers can also poison features while preserving maliciousness, and our experiments verify that the resulting fake variants with poisoned features impaired discriminative classifiers and succeed in lowering the detection score in a test environment. Other main findings are as follows:
\begin{itemize}
\item We observe that different feature categories have different impacts on crafted camouflage samples. The effect rate of API call leads to greater perturbation than permission.

\item We emulate the feature extraction for all types of features that \textsc{Drebin} used and find that \textsc{Drebin}-used feature extraction is substantially more computationally complex and does not necessarily boost the accuracy.

\item We find that in the data-imbalanced (benign-malicious ratio) environment, the accuracy of \textsc{KuafuDet} gradually degrades as we put in more benign applications, but the accuracy still remains relatively high.

\item We find that similarity-based filtering analysis and re-labeling have an excellent performance to handle against adversarial attacks.
\end{itemize}

To the best of our knowledge, this is the first paper to accommodate a newly designed two-phase adversarial machine learning mechanism into mobile malware detection to limit the possibility of mimicry and poisoning attacks, and further propose a learning enhancing system through adversarial detection of Android malware.

The rest of the paper is organized as follows. Section~\ref{problem definition} defines the research problem. Section~\ref{mc} presents the motivations and challenges. Section~\ref{ov} provides the system overview followed by the implementation shown in Section~\ref{tad}. Section~\ref{experimentevaluation} summarizes experimental evaluation. Section~\ref{discussion} discusses limitations. Section~\ref{related} surveys related work. Finally, Section~\ref{conclusion} concludes the paper.

\subsection{A Note on Ethics}
In this paper, we are very aware of the potential impact on malicious apps disclosure or exploited by other malicious third parties. In particular, in order to illustrate this methodology, the collection of mobile malware used and crafted was strictly followed by the Privacy Policy of the Pwnzen Infotech Inc., and conformed to the non-disclosure agreement (NDA) of the Pwnzen Infotech Inc. Furthermore, to facilitate research on mobile malware detection, we make the malicious Android applications (except those from Pwnzen Infotech Inc.) used in the paper publicly available to other researchers under \url{http://nsec.sjtu.edu.cn/kuafuDet/kuafuDet.html}, but no attempt was made to provide data from Pwnzen Infotech Inc. for people outside of our research group because of intellectual property. Only Pwnzen Infotech Inc. authorized employees, using internal computer systems from Pwnzen Infotech Inc., can have access to the dataset. Finally, we informed the team of the Pwnzen Infotech Inc. of the potential newly-discovered malicious apps in order to help Pwnzen Infotech Inc. improve the quality of its products and services. We believe this study performs an important public service, as it shows that even state-of-the-art anti-virus systems are somehow futile. Our ultimate goal is to inform developers and users of such potential poisoning attack, so that more comprehensive countermeasures can be taken in the future.

\section{Problem Definition: Adversarial Machine Learning}\label{problem definition}
We denote a sample set by $\{(x_i, y_i) \in (\mathcal{X}, \mathcal{Y})\}^{n}_{i=1}$, where $x_i$ is the $i$th malware sample vector of which each component exhibits the selected feature; if $x_i$ has the $j$th component, then $x_{ij}=1$; otherwise $x_{ij}=0$. $y_i \in \{0,1\}$, $n$ is the total number of malware samples, and $\mathcal{X} \subseteq {\{0,1\}}^{m}$ is a $m$-dimensional feature space. In this paper, we consider binary classifiers with only two output classes where the attacker crafts malware dataset to evade detection and hence achieves his goals. The attacker tries to move away malware dataset $y_i=1$ in any direction by adding a non-zero displacement feature vector $\delta_i$ to $x_i|_{{y_i}=1}$. For example, attackers may add good attributes to mobile malware to evade binary classifiers. We note that attackers will not be able to modify legitimate benign applications since an honest author has no interest in having his benign application classified as malware. Hence, crafting an adversarial sample ${x^{*}}$, misclassified by the function $f$ (where $f: x\rightarrow y={f}(x)$), from a benign sample $x$ can be formalized as the following problem~\citep{szegedy2013intriguing}:
\begin{equation}
\label{equation1}
{x^{*}}={x} + \delta _{x}\quad \text{s.t.} \quad f({x}+\delta_x)\neq f({x}),
\end{equation}
where $\delta _{_{x}}={x}$ is the minimal perturbation yielding misclassification and $f$ can be the corresponding softmax function.

The goal of adversarial sample crafting in malware detection is to mislead the detection system, causing the classification for a particular application to change according to the attackers wishes. In this paper, we only focus on the \textit{poisoning attack} that results in malicious behavior being misclassified as benign (false negatives), because Inter-Component Communication (ICC) analysis is used to reduce false positives~\citep{li2015iccta,octeau2016combining}. We also assume the attacker has full access to the classifier used, and can inject as many variants' features as possible at will to the given classifier. For this reason, following by the Equation~\eqref{equation1}, we further denote $x_{ij}^{\max}$ ($=1$) and $x_{ij}^{\min}$ ($=0$) as the maximum and the minimum values that the $j$th feature of the $i$th sample can take. Then a poisoning attack can be characterized in the following:
\begin{equation}
\begin{aligned}
\label{bounded}
C_f(x_{ij}^{\min}-x_{ij}) \leq \delta_{ij} \leq C_f(x_{ij}^{\max}-x_{ij}), \quad \forall j \in [1,m],\\
= C_f(0-x_{ij}) \leq \delta_{ij} \leq C_f(1-x_{ij}), \quad \forall j \in [1,m],
\end{aligned}
\end{equation}
where $C_f \in [0,1]$ controls the aggressiveness of attacks. $C_f=0$ indicates no attacks, while $C_f=1$ indicates the most aggressive attacks. To test the ramifications of causative attacks and clearly elaborate the challenges, we develop an adversarial model with three types of attackers with the corresponding $C_f$ values.

{\bf \noindent Weak attacker ($C_f=0.33$).} Our weak attacker is not aware of the statistical properties of the training features or labels at all. This attacker simply fakes additional labels with random binary features to poison the training dataset.
	
{\bf \noindent Strong attacker ($C_f = 0.67$).} Our strong attacker is aware of the features we use for training and can has access to our ground-truth dataset (which comes from public sources). This attacker can manipulate partial features in the training data. However, this attacker is resource constrained and cannot manipulate any mobile application statistics which would require more time. The strong attacker crafts features by randomly selecting public available Android malware and then faking additional labels, so that the partial training labels can become nearly identical.
	
{\bf \noindent Sophisticated attacker ($C_f = 1$).} Our strongest attacker, named sophisticated attacker, has full knowledge of our training feature set. Additionally, this attacker has sufficient time and economic resources to create arbitrary mobile application statistics. Therefore, the sophisticated attacker can fully manipulate almost all training features, which creates scenarios where relatively benign mobile applications and real-world malicious mobile applications appear to have nearly identical attributes at the training phase.

To do this, we adopt the adversarial crafting algorithm~\citep{papernot2016limitations} based on the Jacobian matrix $$J_{f}=\frac{\partial f\left (\mathcal{X} \right )}{\partial \mathcal{X}}=\left [ \frac{\partial f_{i}\left ({x} \right)}{\partial {x}_{j}} \right ]_{i\in \{0,1\}, j\in \left [1,m \right]}$$ where $f_{0}(x)$ outputs $x$ is benign and $f_{1}(x)$ outputs $x$ is malicious, with $f_{0}\left ( x \right )+f_{1}\left ( x\right )=1$. To craft an adversarial sample, we use $73$ benign features and $102$ malicious features detailed in Section~\ref{featuresec}. We then take mainly two steps. In the first step, we compute the gradient of $f$ with respect to $x$ to estimate the direction in which a perturbation in $x$ would change $f$'s output. In the second step, we choose a perturbation $\delta$ of $x$ with maximal positive gradient into our target class $\mathcal{Y}_{y_i=0|y_i=1}$ denoted $y^{\prime}$, and we customize the adversarial crafting algorithm~\citep{papernot2016limitations} according to our adversarial model with three types of attackers ($C_f$), to indicate the probability (Equation~\eqref{bounded}) of adding a specific feature. After computing the gradient, we iteratively choose a target feature of which the gradient is the largest for our target class and then update its value in $x$  to obtain our new input vector. We then re-update the gradient and repeat this process until either (\rmnum{1}) we reach the bounded allowed changes (loop bound) or (\rmnum{2}) we successfully achieve a misclassification.

We didn't attempt to formalize the malware detection as an optimization problem, saying to maximize accuracy at the lowest resource cost with minimal adversarial perturbations, because simply using optimization may not be aligned with how we semantically understand the human malicious behaviors and motivations. For future research, we wish to generate an exact mapping rule between machine-crafted mobile malware and read-to-use malicious apps in the wild, and ultimately wish to provide a foundation for developing sustainably-secure anti-malware systems in the face of dynamic cyber-maneuvers.

\section{Motivations and Challenges}\label{mc}
In this section, to highlight our contributions, we motivate our malware detection model by incorporating adversarial environment witnessed in the real world and then review several challenges of our work.

\begin{figure*}[!htb]
  \centering
  \includegraphics[width=0.9\textwidth]{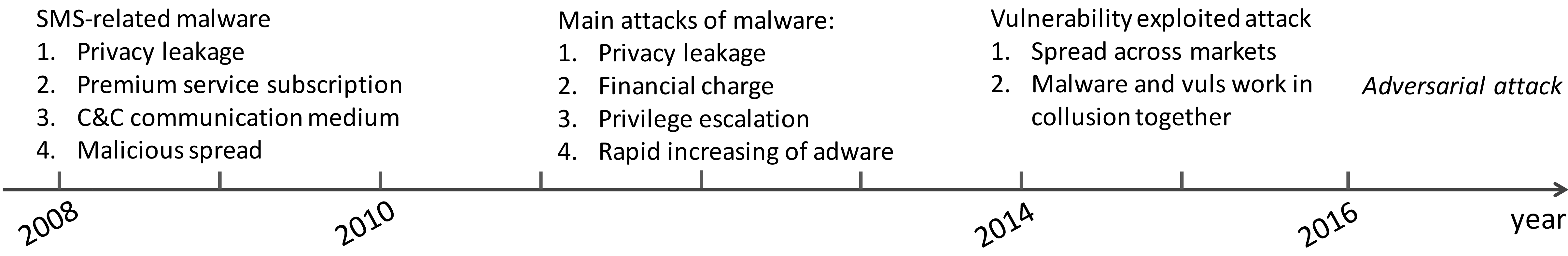}
  \caption{A timeline of Android malware}
  \label{fig:timeline}
\end{figure*}

\subsection{Evolutionary Chain}
\begin{figure}[!htb]
  \centering
  \includegraphics[width=0.45\textwidth]{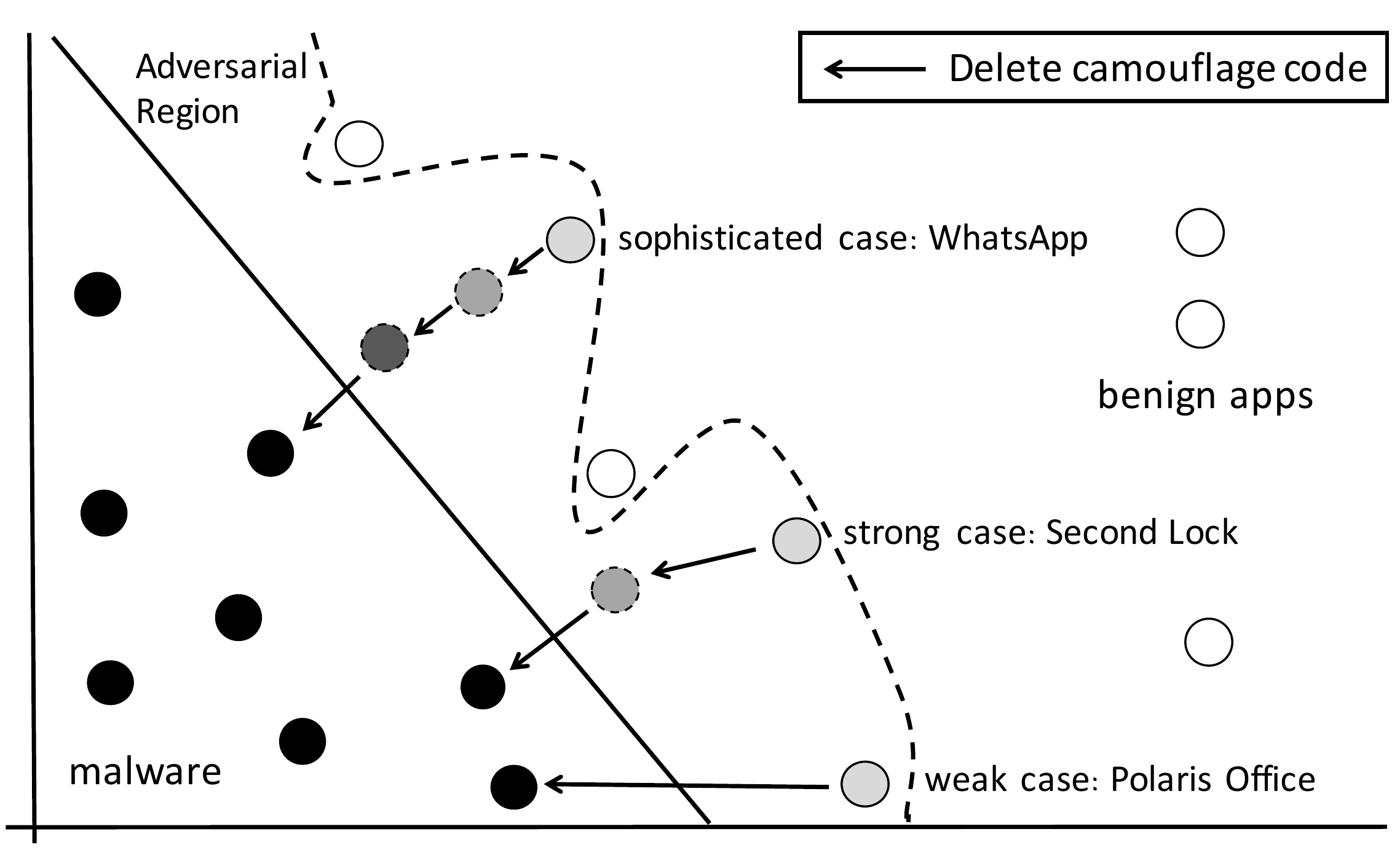}
  \caption{The process from benign to malicious: the black dots refer to malicious apps, the white dots refer to benign apps, and the gray dots are originally malware but misclassified as benign due to the camouflage code injected in the adversarial region. The gray dots turn darker with the deletion of camouflage code and finally turn all black, exposing its malicious nature. The arrows direct toward the transformation process.}
  \label{fig:casestudy}
\end{figure}

Figure~\ref{fig:timeline} describes an evolutionary chain of mobile malware advancements that has been observed along the timeline: ranging from the seed explosion to the recent adversarial attack. Earlier menaces of malware are mainly pertinent to compromised SMS related functions. With every new technology comes abuse, the Android market is no exception. Since 2014, malware samples can be easily exploited by Android vulnerabilities, which heavily drives an arms race for the adversarial detection of mobile malware. As the nature of the attack is shifting from small-scale and low-tech towards large-scale and skilled ones, the additional efforts will have to be directed at taking targeted strategies to detect stubborn malware.

\subsection{Adversarial Samples}
In previous work~\citep{chen2016stormdroid}, a number of malware samples in the dataset are misclassified into benign ones.  We zoom in these misclassified samples and witnessed several real-world cases that reflect adversarial attacks from this dataset. Each observation corresponds to the three types of attackers defined above. 

\begin{figure}[!htb]
  \centering
  \includegraphics[width=0.55\textwidth]{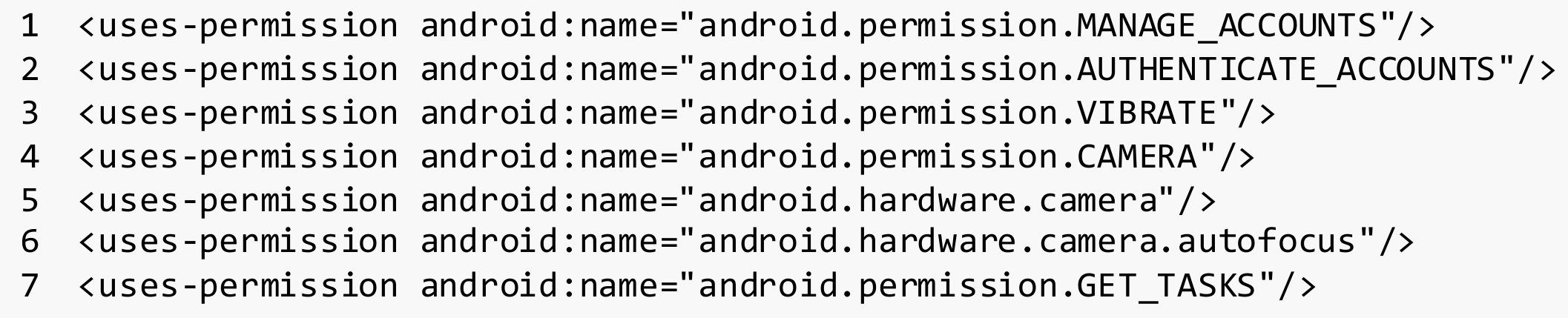}
  \caption{A case of permissions overprivileged}
  \label{case1}
\end{figure}

\vspace{2mm}
\noindent {\bf Weak attacker.} Embedding a good portion of benign code into a malicious app (\textit{e.g.}, manifest attributes and non-logical code in java code). Felt~\emph{et al.}~\citep{felt2011android} show that more than half of Android applications are overprivileged, such as misusing \textit{AndroidManifest.xml} configuration.
We carefully analyze the false negative samples using mapping relations between permissions and API calls, and find that some declared permissions remain unused at the development stage. We can see from Figure~\ref{case1} that part of the code in the \textit{AndroidManifest} file of \textit{Polaris Office} is misclassified as benign. Although \#1 to \#7 permissions are declared, they have not been used at all since the app does not demonstrate corresponding behaviors. 
These permissions extracted as features for training classifiers weakly mislead the classification outcomes. 
As shown in Figure~\ref{fig:casestudy}, we delete redundant code (\#1 -- \#7) that is irrespective of the logical behavior of the sample. After we repeat the classification process, surprisingly the sample is classified as malicious.

\noindent {\bf Strong attacker.} Hiding a good portion of malicious code into other formats in the application package.
Some alternative malicious applications use techniques such as dynamic loading techniques to hide a good portion of malicious code into other formats. Usually these malicious code blocks include sensitive API calls. API calls used by most of machine learning classifiers will dramatically lead to the misclassification. 
For example, \textit{SecondLock} behaves as same as ``egdata'' does. Some malicious code blocks or executable files are hidden in the files of other formats, such as \texttt{jar}, \texttt{so}, \texttt{jpg}, and \texttt{data}. These files contain sensitive API calls, such as \texttt{DownloadManager.enqueue} and \texttt{DownloadManager.query}, which are hidden by malware. The ``assets'' folder of \textit{SecondLock} contains a \texttt{png} format file, which is not a standard \texttt{png} file and can be dynamically loaded. It can prevent the application from updating or automatically downloading other malicious applications. As shown in Figure~\ref{fig:casestudy}, we show two steps to correct the classification results. In step one, we remove five unused permissions, the result of classification moves toward the hyperplane, though the final result remains benign. In step two, we add some sensitive API calls that are hidden in the \texttt{png} file (\textit{e.g.}, \texttt{DownloadManager.enqueue} and \texttt{DownloadManager.query}). This particular sample is finally classified as malicious.

\begin{figure}[!htb]
  \centering
  \includegraphics[width=0.55\textwidth]{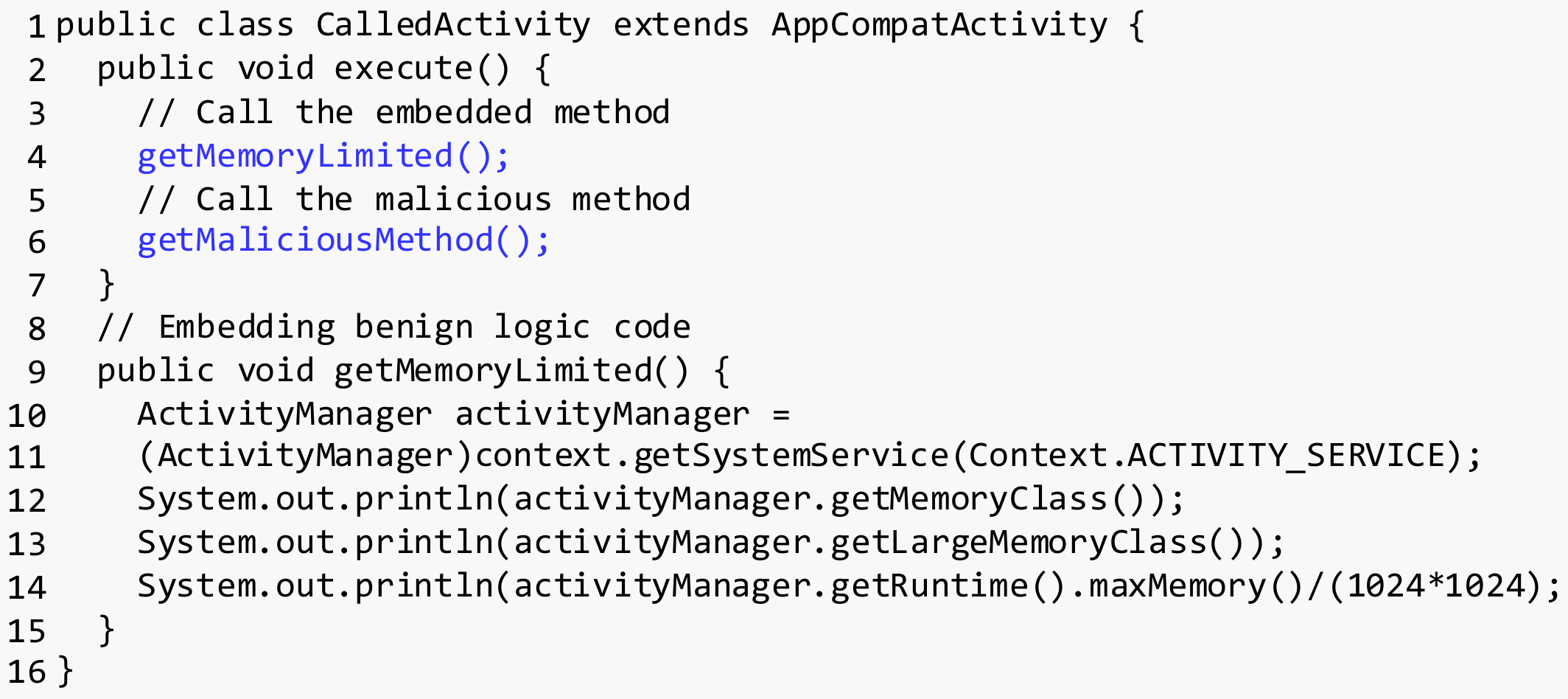}
  \caption{A case of Activity embedded with benign API calls}
  \label{fig:case3}
\end{figure}

\vspace{2mm}
\noindent {\bf Sophisticated attacker.} Embedding benign logic code in java code and dynamic code loading with reflection.
A few malicious applications add benign logic code in source code. Benign logic code can be executed without any effects on malicious behaviors, which is used to obscure the feature extraction process to clone benign applications. This is similar to ``testing code.'' By embedding benign logic code, the sophisticated attacker can add any code blocks or any combination of various techniques to mislead the machine learning classifiers, making the classifiers less robust. 
Specifically, after the construction of Activity transition relations for \textit{WhatsApp}, a repackaged malware from the third party rather than the official version, we find there exist embedded activities, standing alone with some methods, such as neither being affected by any other activities nor shown up in the system logic. Figure~\ref{fig:case3} shows a code segment of \textit{WhatsApp} with an embedded activity, \texttt{getMemoryLimited()}method initiates a system call to \texttt{getMemoryClass()}, \texttt{getLargeMemoryClass()}, etc. However, these system calls extracted as features for training cannot reflect system logic of the sample, which seriously misleads the classification outcomes. We therefore remove such embedded benign logic code and retrain the classifiers. As shown in Figure~\ref{fig:casestudy}, once we delete such code step by step, the malware sample is exposed. Figure~\ref{fig:casestudy} also exemplifies that in the adversarial environment, the attack process is changeable and dynamic. 

\begin{figure}[!htb]
  \centering
  \includegraphics[width=0.55\textwidth]{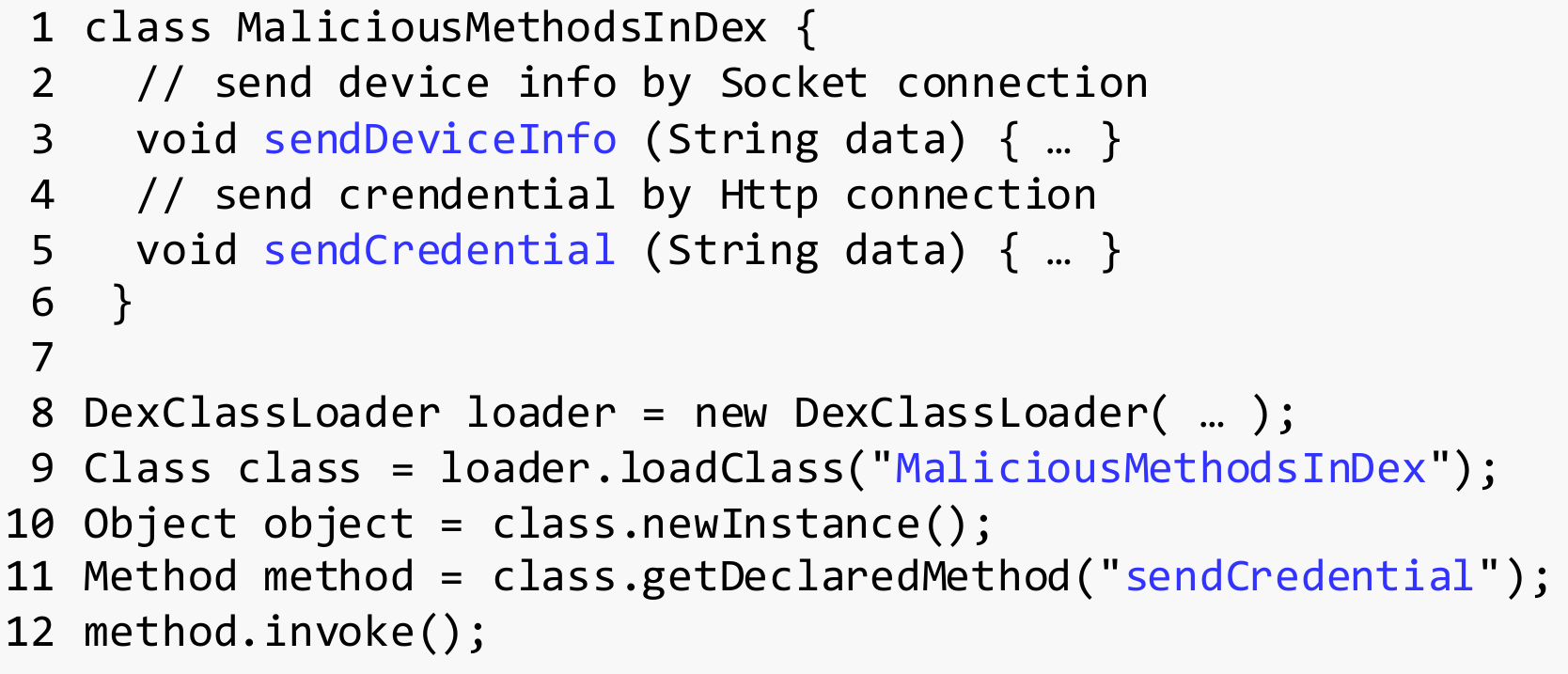}
  \caption{An loading example of using reflection (demonstrated as dynamic pseudo code)}
  \label{fig:dynamic}
\end{figure}

The alternative approach that one sophisticated attacker may take is through dynamic code loading. Dynamic code loading usually utilizes reflection mechanism to modify the runtime behavior of applications. It provides ability for sophisticated attackers to add malicious behaviors (malicious features) without having to change the original application, hence mislead the machine learning classifiers. For example,  as depicted in Figure~\ref{fig:dynamic}, sophisticated attackers can incorporate two malicious methods in the class \texttt{MaliciousMethodsInDex} (lines \#1 -- \#6) -- \texttt{sendDeviceInfo} and \texttt{sendCredential}, and instantiate them via \textsc{DexClassLoader} at runtime. Since the malicious codes are loaded at runtime, and not part of the application source, it is challenge to classify them as malware through machine learning classifiers.

\subsection{Challenges}\label{challenges}
{\bf \noindent Class imbalance.} We aim to train a classifier that produces binary predictions: each mobile application is classified as either benign or malicious. If there are significantly more malicious applications in one class than in the other class, this biases the output of supervised machine learning algorithms. Prior research treats it simply by using 49 different malware families~\citep{zhou2012dissecting}. In consequence, our dataset exhibits a modest class imbalance:  We first define 217 Android malware families and then classify them into eight categories, such as Expense, Fraud, Payment, Privacy, Remote, Rogue, Spread, and System (see Table~\ref{fig:category}), to present a systematic characterization of existing Android malware. The eight malware categories represent almost all coverage of existing Android malicious behaviors. We observe that it is common that many malware families belong to multiple categories in parallel. This phenomenon indicates that malicious family is not limited to a single malicious behavior. As shown in Table~\ref{fig:category}, Privacy and Fraud occupy the highest proportions among all. Therefore, we are able to find the reasonable distribution of our real-world malware during our data acquisition. As the ratio of malware to benign apps in the real-world is highly imbalanced, this class imbalance usually represents a significant challenge for reducing false negatives.

\begin{table}[!htb]\footnotesize
\caption{Malware category}
 \label{fig:category}		
  \begin{minipage}{\columnwidth}
 \begin{center}
 \begin{tabular}{|l|c|}
 \hline {\bf  Malware category}  & {\bf Percentage}  \\
 \hline
 \hline  Privacy	  	& 47.1\% \\
  \hline Fraud  		& 33.9\% \\
  \hline Rogue 	   	& 20.1\%  \\
  \hline Spread  	        	& 20.1\% \\
  \hline System 	        	& 11.1\%  \\
  \hline Remote         	& 10.6\% \\
  \hline Expense       	& 10.1\% \\
  \hline Payment      	& 6.3\%  \\
  \hline
\end{tabular}
\end{center}
 \bf Note: We take a look at malware category with miscellaneous datasets. We find that many Android malware families belong to multiple malware categories, thereby the sum of percentages is greater than 100\%.
\end{minipage}
\end{table}

{\bf \noindent Quality of ground-truth dataset.} Prior work on malware dataset focused on validating their approaches by using a small out-dated dataset. These predictors can be used as ground truth for training high-performance classifiers. In contrast, there is no comprehensive dataset of malware that are available in the real world.  We employ as ground truth the set of malware from five different platforms. In particular, the set obtained from Pwnzen Infotech Inc. is the most recent malware in the real world. However, we acknowledge this dataset does not cover all platforms uniformly.

\section{System Overview}\label{ov}
In this section, we provide a high-level overview for our system design.
\subsection{Key Ideas}
In putting our approach in practice for massive-scale detection, we aim to achieve two important goals: one is to identify and filter the suspicious false negatives (\textit{i.e.}, the malicious applications that are camouflaged), the other is to achieve accuracy and scalability at the same time. To achieve these two goals, we propose two key techniques: similarity-based filtering and two-phase iterative adversarial detection, as shown below.

\noindent  \textbf{(\rmnum{1}) An automated similarity-based approach to filter suspicious false negatives.} In general, attackers have two characteristics: first, attackers may acquire dual intrinsic attributes of applications, thus we assume the suspicious ones are the ones with both strong reflection on malicious features and benign features; second, in our threat model, attackers may have certain level of knowledge of training set, thus we decouple the similarity metrics from machine learning classifiers. Based on these two characteristics, if the trained classifier could incorporate the similarities across the applications in the training set to lead to a further fine-grained detection, the learning system would be periodically enhanced with these newly discovered malware and suspicious false negatives.

\noindent \textbf{(\rmnum{2}) A two-phase iterative adversarial detection approach to achieve accuracy and scalability.} False negatives can be reduced based on our understanding of an attacker's threat level. Thus self-adaptive learning (SAL) scheme, where the suspicious false negatives, as well as the identified malicious applications, are fed back to the training process for a desired trade-off between accuracy and scalability.

\subsection{Overall Architecture of KuafuDet}
\begin{figure*}[!htb]
  \centering
  \includegraphics[width=0.7\textwidth]{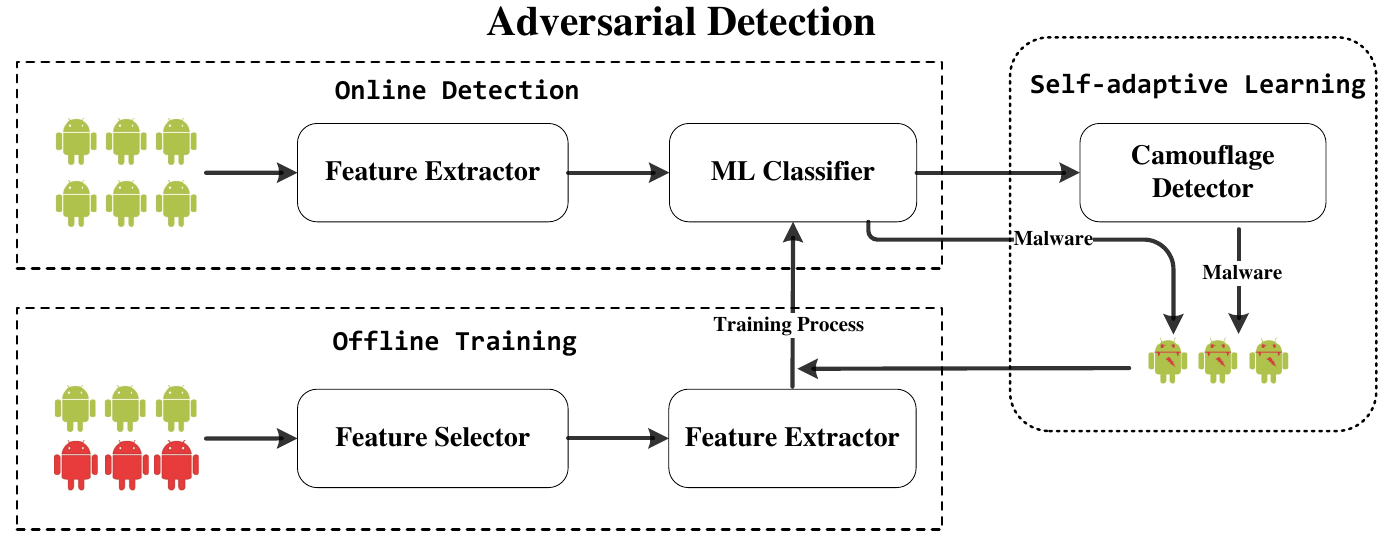}
  \caption{The \textsc{KuafuDet} framework through adversarial detection}
  \label{fig:architecture}
\end{figure*}

The overall architecture of \textsc{KuafuDet} is shown in Figure~\ref{fig:architecture}, which is comprised of two intertwined phases. In the {\bf Training Model} phase, \textsc{KuafuDet} extracts features from labeled applications based on our combined set of contributing features, trains classification model offline, and updates classifiers at a certain interval of time; In the {\bf  Online Detection} phase, \textsc{KuafuDet} classifies large sets of online Android applications (from multiple online Android markets) into different categories, benign and malicious; Meanwhile, \textsc{KuafuDet}, through a {\bf  Self-adaptive Learning} scheme, discovers new information from both the identified malware and the filtered suspicious false negatives from {\bf Camouflage Detector}, and incorporates into {\bf Training} to stabilize the detection accuracy.

\section{System Design}\label{tad}
In retrospect, the quality of discriminative classifier is the key to the accuracy of malware detection. On one hand, when the trained classifier is trained once and used for all time, it is not able to corresponding to the new malware. On the other hand, aggressive attackers may obfuscate their representations in terms of contributing features to impair discriminative classifiers. Thus it might lead to high false negatives that the malicious applications evade detection. In order to perform accurate and scalable adversarial detection, our proposed adversarial detection approach contains two phases, training and detection, intertwined by the self-adaptive learning (SAL) scheme. In particular, we conduct our similarity-based analysis in Camouflage Detector to filter the suspicious false negatives.

The implementation of \textsc{KuafuDet} involves the following steps:
\begin{enumerate}
\item In the feature selection stage, we decompile APKs to generate \textit{Smali} code via Apktool,\footnote{\url{http://ibotpeaches.github.io/Apktool/}} we extract 195 out of 564 features using manual pruning along with information gain validated.

\item In the training stage, we use different machine learning classifiers, such as Support Vector Machine (SVM), Random Forest (RF), and $K$-Nearest Neighbor (KNN), based on 195 dimensional features we selected.

\item In the camouflage detection stage, we perform similarity-based filtering to identify the false negatives that are the camouflaged malicious applications.
\end{enumerate}

\begin{table*}[!htb]\scriptsize \renewcommand{\arraystretch}{1.2}
\caption{Features}
\label{fig:features}
 \begin{center}
\begin{tabular}{|c|c|c|c|}
\hline
\multicolumn{3}{|c|}{{\bf Syntax Features}} & {\bf Semantic Features}\\ \hline
\hline
{\bf Permission} & {\bf Intent \& Hardware}& {\bf API Call}& {\bf Sequence }\\ \hline
READ\_PHONE\_STATE & INTENT.ACTION.DELETE & URL.openConnection & (chmod 777, Runtime, getRuntime, exec)\\ \hline
WRITE\_SMS & INTENT.ACTION.GET\_CONTENT & TelephonyManager.getDeviceId & (getDeviceId, URL, openConnection)\\ \hline
INSTALL\_PACKAGES & HARDWARE.TOUCHSCREEN & PackageManager.checkPermission &  (DownloadManager, Uri, Request, enqueue)\\ \hline
\ldots \ldots& \ldots \ldots& \ldots \ldots & \ldots \ldots\\ \hline
\end{tabular}
\end{center}
\end{table*}

\subsection{Feature Selector}\label{featuresec}
The features considered in this study are classified into two categories: syntax features ({\footnotesize {$\mathcal{S}^{\{PERM, \, INT, \, H, \, API\}}$}}) and semantic features ({\footnotesize {${\mathcal{S}^{\prime}}^{\{Sequence\}}$}}).

\subsubsection{Syntax features} 
Through closely examining more than 250,000 applications from various sources (breakdowns shown in Section~\ref{dataset}), we notice that the malicious applications tend to have drastically different permissions, intents, hardwares and API calls, which supports the assumption that malicious applications are distinguishable from benign ones. To facilitate reading, we show a coarse-grained description of syntax features used in this paper.

\begin{itemize}
\item Permission ({\footnotesize {$\mathcal{S}^{\{PERM\}}$}}): Each APK has an AndroidManifest file in its root directory, which is an essential profile including information about the application. Android OS must process this profile before it runs any of installation. The profile file declares which permissions the application must have in order to access protected parts of the API and interact with other applications. It also declares the permissions that others are required to have in order to interact with the application's components.

\item Intent ({\footnotesize {$\mathcal{S}^{\{INT\}}$}}): Communication between different components is mainly through intent, which can be regarded as the ``medium'' where information about massive asynchronous data exchange and calls to different components is shared between different components and applications.

\item Hardware ({\footnotesize {$\mathcal{S}^{\{H\}}$}}): Features about requesting access to specific hardware of the smartphone should be declared in the manifest file, such as NFC and GPS, since the combination of such hardware modules may have harmful impact on the phone.

\item API Call ({\footnotesize {$\mathcal{S}^{\{API\}}$}}): Android API calls monitoring, based on the reverse engineering, can monitor those API calls, such as sending SMS, accessing user location, and device ID. The Android platform provides a framework API that applications can be used to interact with the underlying Android OS. The framework API consists of a core set of packages and classes. Most applications use a fairly large number of API calls.
\end{itemize}

Here, we use statistical metrics-driven manual pruning~\citep{chen2016stormdroid} with information gain to cross-check the feature selection. Although information gain facilitates the automatic feature selection, it ignores the class information and distribution of the features. When these features are used to detect malware, the performance would drop down dramatically. For example, \texttt{ READ\_INPUT\_STATE} (resp. \texttt{ ACTION.SET\_WALLPAPER})  corresponds to Permission (resp. Intent) exhibits a high information gain than many others but it relates to only a small subset of malware. Such highly specific features are undesirable for classification. In summary, we generate 175 types of syntax features. 

\begin{equation*}
\begin{aligned}
\label{syntax}
& \sum \overbrace{\underbrace{\# \bigcap_{}^{  } {\mathcal{S}^{\{PERM\}}}}_{\bf 61} + \underbrace{\# \bigcap_{}^{} {\mathcal{S}^{\{INT\}}}}_{\bf 12} +   \underbrace{\# \bigcap_{}^{} {\mathcal{S}^{\{H\}}}}_{\bf 5} + \underbrace{\# \bigcap_{}^{} {\mathcal{S}^{\{API\}}}}_{\bf 97}}^{{\mathcal{S}^{\{PERM, \, INT, \, H, \, API\}}}}. \\
\end{aligned}
\end{equation*}

\subsubsection{Semantic features}
The semantic feature ({\footnotesize {${\mathcal{S}^{\prime}}^{\{Sequence\}}$}}) represents malicious behaviors that occur sequentially, which are extracted via static analysis. For instance, the sequence \texttt{(DownloadManager, Uri, Request, enqueue)} in Table~\ref{fig:features} indicates that a download request requires to follow a certain order: construct a  request object and transfer the URL of the file to \textit{enqueue} method to finish the download process.
These sequences can characterize some interesting malicious behaviors that cannot be captured by the syntax features and can reflect the malicious behaviors more explicitly for a large number of apps, with the purpose of training classifiers. We de facto take several sensitive behaviors into consideration, such as ``Send SMS,'' ``Request for chmod,'' ``Uninstall application,'' ``Get location,'' ``Get wifi info,'' and ``Start http connection.'' We then extract the sequence of key strings that reflect interesting malicious behaviors. For example, Requesting for chmod is described as the sequence of ``chmod 777,'' Runtime, getRuntime, and exec, we define 20 types of semantic features for detecting malware. By generating Android malware, these semantic features can be extended. 

\begin{equation*}
\begin{aligned}
\label{semantic}
& \sum {\underbrace{\# \bigcap_{}^{} {\mathcal{S}^{\prime}}^{\{Sequence\}}}_{\bf 20}}. \\
\end{aligned}
\end{equation*}

To characterize each of the applications using static analysis, we generate a final set of 195 out of 564 types of features, as partially shown in Table~\ref{fig:features}. In summary, we use 195-dimensional feature vectors for the study (breakdowns shown in Table~\ref{fig:features_whole}).

\subsubsection{Justifying Feature Selection}

In \textsc{Drebin}, the feature set contains thousands of arbitrary strings that appear in the manifest file or in the disassembled code of the app chosen by developers. In particular, 
\textsc{Drebin} extracts eight types of features: hardware components, requested permissions, App components, filtered intents, restricted API calls, used permissions, suspicious API calls, network addresses. These massive features chosen from his dataset act as noises, misleading the classifier. We readily emulate the feature extraction for all types of features that \textsc{Drebin} used since \textsc{Drebin} authors do provide the feature vectors of their own dataset for evaluation by other researchers. As shown in Table~\ref{fig:result111}, we use \textsc{Drebin} dataset~\citep{arp2014drebin}, massive \textsc{Drebin}-used features, and simulate his algorithm. We conclude that extracting \textsc{Drebin}-used features was a substantially more computationally complex process than our feature selection due to the sheer number of features extracted. In fact, these features do not necessarily boost the accuracy. Our approach also validates that 195 types of features used in \textsc{KuafuDet} are proper and will not trigger the classical curse of dimensionality.

\begin{table}[!htb]\footnotesize
\caption{Comparison of consequential features (using \textsc{Drebin} dataset)}
\label{fig:result111}
 \begin{minipage}{\columnwidth}
 \begin{center}
 \begin{tabular}{|l|c|}
 \hline {\bf  Detection Tool within Features Used}  & {\bf Accuracy}\\
 \hline
 \hline 195-dimensional features used in \textsc{KuafuDet} 	  		&  96.55\%\\
 \hline 564-dimensional features used in \textsc{KuafuDet}  			& 95.80\% \\
 \hline 5,000 features used in \textsc{Drebin} 		   				& 94.05\%\\
 \hline 500,000 features used in \textsc{Drebin} 			& 93.90\%\\
 \hline
\end{tabular}
\end{center}
\end{minipage}
\end{table}

\subsection{Machine Learning Classifiers}
With these 195-dimensional features that result from our feature selector, we utilize a number of popular algorithms widely used in security contexts, including Support Vector Machine (SVM), Random Forest (RF), and $K$-Nearest Neighbor (KNN). We leverage existing implementations of these algorithms in WEKA~\citep{hall2009weka}. In particular, SVMs seek to determine the maximum margin hyperplane to separate the classes of malicious and benign applications. When a hyperplane cannot perfectly separate the binary class samples based on the features we fed in, we then tune the parameters such as regularization penalty and non-negative slack variables. We also perform multiple rounds of stratified random sampling due to the data imbalance as stated in Section~\ref{challenges}. Random Forest (RF) and $K$-Nearest Neighbor (KNN) are also tuned in an analogous manner.

\subsection{Camouflage Detector}
To further discover camouflage in malware, we manually pick a fair number of applications from the farthest very benign outcomes and very malicious outcomes from the classification hyperplane, respectively. In particular, these three machine learning algorithms use the corresponding \textit{distance} to classification hyperplane. And those hand-picked applications are the most benign and most malicious predictions and would be updated along with training set updating. We assume these applications have not been poisoned by any malicious third parties. We then measure the similarity between the training set and the selected most benign applications, and vice versa the selected most malicious applications after classification. By further tuning the similarity threshold, we relabel the camouflage malware of the training set as malicious samples to make the classifier robust. Moreover, based on similarity analysis, we are able to identify the camouflage malware from false negatives. Our similarity-based approach is based on extracted robust features and those non-bypassed samples that are farthest from the hyperplane.

\begin{figure}[!htb]
\begin{minipage}[t]{0.5\linewidth}
\centering
\includegraphics[width=1.7in]{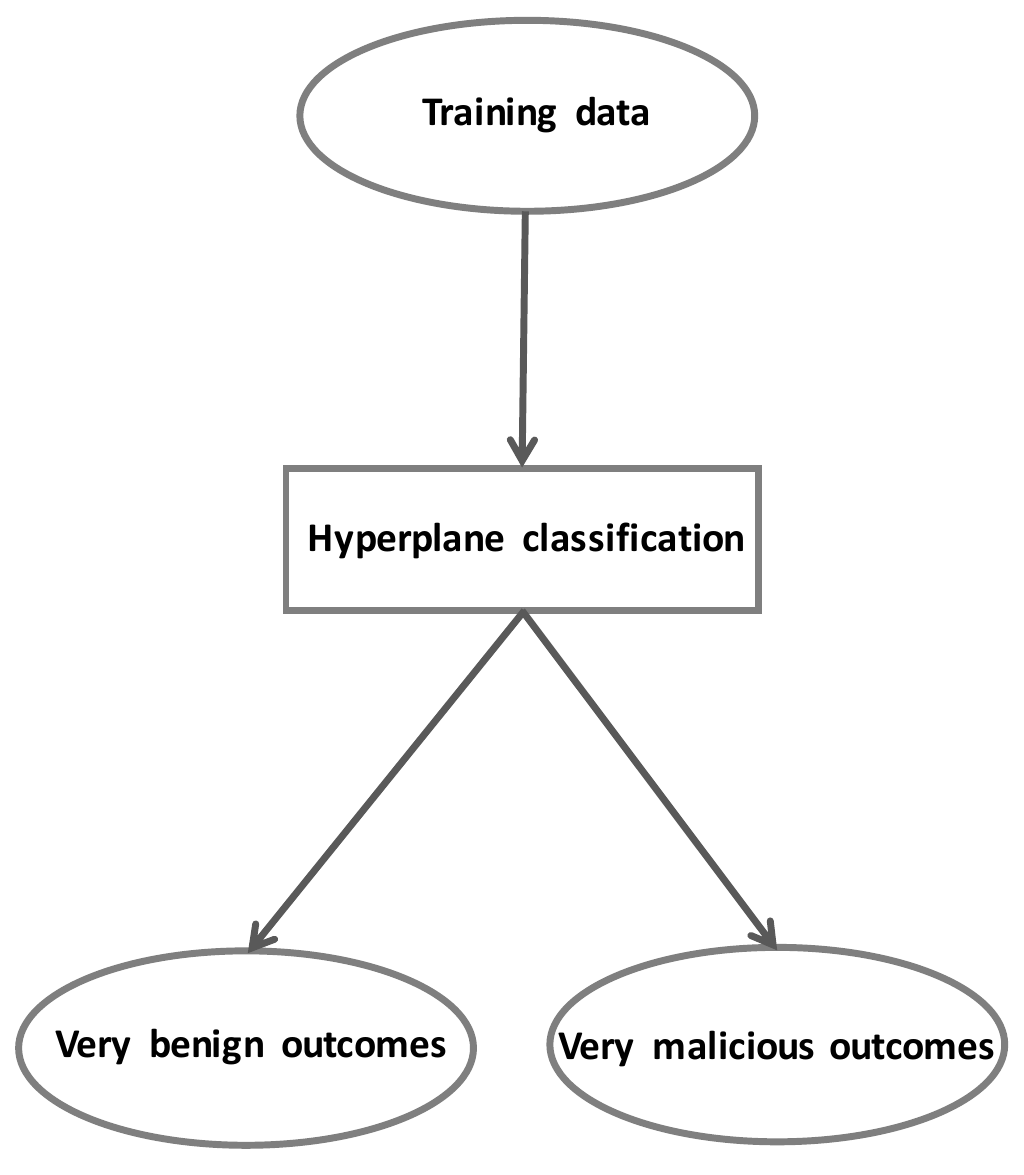}
\caption{Choosing very benign and malicious outcomes}
\label{fig:vary}
\end{minipage}%
\begin{minipage}[t]{0.5\linewidth}
\centering
\includegraphics[width=2in]{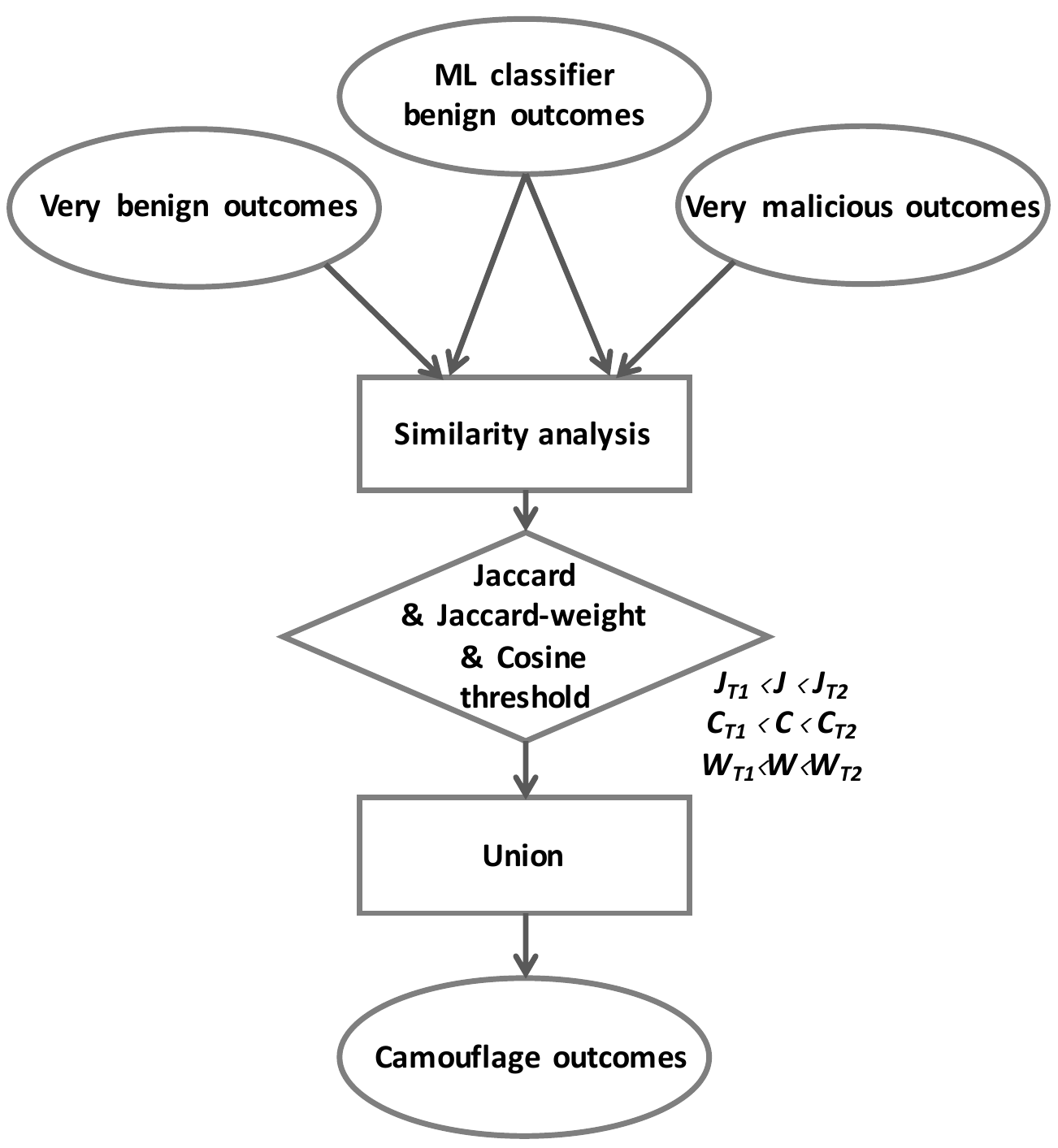}
\caption{Similarity analysis}
\label{fig:simiarity}
\end{minipage}
\end{figure}

\subsubsection{Measuring Similarity}
We use Jaccard index, Jaccard-weight similarity, and Cosine similarity to measure the similarity of applications. The similarity indices are characterized by two vectors $A$ and $B$, where $A$ represents the feature vectors of the applications in the training set and $B$ represents the feature vectors of hand-picked (the most benign or the most malicious) applications. The similarity indices used in this paper are the following:

\noindent \textbf{Jaccard index}: The Jaccard index, denoted by $J(\cdot)$, is defined as the size of the intersection divided by the size of the union of the sample sets $A$ and $B$:

        \begin{displaymath}
       J(A,B) = {{|A \cap B|}\over{|A \cup B|}} = {{|A \cap B|}\over{|A| + |B| - |A \cap B|}}.
        \end{displaymath}

Jaccard index is not accurate enough because it does not reflect the actual differences of frequencies. For example, an API is used 10 times and 100 times respectively in two  applications, but the Jaccard distance simply treats them equally.

\vspace{2mm}

\noindent \textbf{Jaccard-weight similarity}: The Jaccard-weight similarity is defined as follows, which is computed by two steps.

Step 1. The weight of each component of the feature vector $J_f$ is defined as the percentage that the number of apps which exhibit that feature over the total number of apps.

 Step 2. For any two app $a \in A$ and $b \in B$, we consider if both the $k$th component of feature vectors are non-zero, that is $a^{k}=b^{k}=1$, where $a^{k}$ and $b^{k}$ denote the ${k}$th component of feature vectors of apps $a$ and $b$, respectively. We collapse the Jaccard-weight similarity $W(a,b)$ as follows:

\begin{displaymath}
  	 W(a, b)= \frac{\sum_{k=1}^{n} J_{f}^{k} \mathds{1}(a^{k}=b^{k}=1)} {\sum_{k=1}^{n} J_{f}^{k}},
	\end{displaymath}
where $n=175$ is the dimension of features and $\mathds{1}(\cdot)$ is the indicator function.

\vspace{2mm}

\noindent \textbf{Cosine similarity}: The cosine similarity, denoted by $\cos(\theta)$, is defined using a dot product of the two vectors $A$ and $B$ divided by the product of their magnitudes as:
        \begin{displaymath}
        \cos(\theta) = \frac{A \cdot B}{\|A\| \|B\|}.
        \end{displaymath}

We use Jaccard index ($J$), Jaccard-weight similarity ($W$),  and Cosine similarity ($C$) to measure the similarity of applications. If the similarity between two applications exceeds a certain threshold, the application will be selected as a malware candidate and fed back to the training process for further fine-grained detection. We want to select as many malware candidates as possible for periodically retraining the classifiers. To be specific, a low threshold likely leads to high false negatives, while a high threshold leads to low false negatives. As shown in Figure~\ref{fig:simiarity}, during our experiments, we empirically choose parameters $J_{T_1} < J < J_{T_2}$,  $W_{T_1} <  W < W_{T_2}$, and $C_{T_1} < C < C_{T_2}$ as corresponding thresholds and then take the union of three outcomes for picking the camouflage malware. From an attacker's perspective, in order to evade the detection, the fraction of two sets $A$ and $B$ must be below a given threshold $0<p<1$ for Jaccard index: ${|A \cap B|} \leq p\times|A|$ and ${|A \cap B|} \leq p\times|B|$. An optimal attack strategy is to schedule a group of accounts according to the set of such action sets $A$ or $B$ that has the maximum cardinality so as to minimize the probability that two accounts are caught. But finding $A$ with the maximum cardinality remains an open problem in intersection set theory~\citep{brunk2009intersection}, which poses a limitation to the attacker.

\section{Experimental Evaluation}\label{experimentevaluation}
We evaluate \textsc{KuafuDet} with applications downloaded from different popular third-party Android markets, as well as in real industrial environments such as Pwnzen Infotech Inc. The goals are to evaluate our system in aspects of: (\rmnum{1}) the robustness of our detection under three attacks; (\rmnum{2}) the capabilities of accurately detecting malicious applications; (\rmnum{3}) the efficiency and scalability of real-time analysis, and adaptability to new Android malware; and (\rmnum{4}) the capabilities of detecting coverage.

\begin{table*}[!htb]\footnotesize
\caption{Datasets for adversarial detection of Android malware}
   \label{fig:dataset}
 \begin{center}
\begin{tabular}{|l|c|c|c|c|c|c|}
\hline
\multicolumn{2}{|l|}{\bf Source} & {\bf Universal} & {\bf Data-driven Analysis}  & {\bf Training} & {\bf Test} & {\bf Comparison} \\ \hline
\hline
\multicolumn{2}{|l|}{\multirow{2}{*}{{\bf Benign}}} & \multirow{2}{*}{\bf 242,500} & \multirow{2}{*}{10,000} & \multirow{2}{*}{8,000}  & \multirow{2}{*}{2,400} & \multirow{2}{*}{0} \\
\multicolumn{2}{|l|}{}&  &   &  &   &  \\ \hline
\multirow{4}{*}{{\bf Malicious}}	 	
				   & {\bf Pwnzen Infotech Inc.}                                         & 4,500  & 4,500    	 & 3,500        	    	& 1,000    	& 600    	\\ \cline{2-7}
                                     & {\bf Zhou \textit{et al.}~\citep{zhou2012dissecting}}   & 1,260  & 1,000 	 & 1,000 	 		& 260    	& 150 	\\ \cline{2-7}
                                     & {\bf Arp \textit{et al.}~\citep{arp2014drebin}}		 & 4,300  & 4,200  	 & 3,200 	 		& 700 	& 150 	\\  \cline{2-7}
                                     & {\bf Contagio}   							 & 340    &  300 	         & 300        	 	& 40    	& 100 	\\ \hline
\multicolumn{2}{|l|}{\multirow{2}{*}{{\bf  Apps}}} & \multirow{2}{*}{\bf 252,900} & \multirow{2}{*}{20,000} & \multirow{2}{*}{16,000}  & \multirow{2}{*}{4,000} & \multirow{2}{*}{1,000} \\
\multicolumn{2}{|l|}{}& &  & &  & \\
\hline
\end{tabular}
\end{center}
\end{table*}

\subsection{Experimental Dataset}\label{dataset}
As mentioned earlier, most studies lack a large number of data samples. We fulfill the need by presenting the first large collection of 252,900 Android application samples, including 10,400 malicious samples, which covers the majority of existing to recent ones, as shown in Table~\ref{fig:dataset}. Specifically, these 252,900 APK files we collected consist of 242,500 benign applications that are downloaded from Google Play Store, and the other 10,400 malicious APK files where 1,260 have been validated in \citep{zhou2012dissecting} and the remaining are downloaded from Contagio Mobile Website (340 APKs), Pwnzen Infotech Inc. (4,500 APKs) and \citep{arp2014drebin} (4,300 APKs). Our malicious applications include all varieties of the threats for Android, such as phishing, trojans, spyware, and root exploits. In the following, we randomly select various portions of benign apps and malicious apps (various ratios of \# benign \textit{vs} \# malicious) for different experimental goals. Specifically, we select 1,000 malware as samples out of the set of 10,400 malicious samples and scan them using \textsc{KuafuDet} and other industrial malware detecting tools. For comparison, the 1,000 samples contain both benchmarks before 2014 and the most recent datasets, more than half of which are the most recent malware.

Finally, we measure the efficiency and scalability of \textsc{KuafuDet} performance, and perform the entire process of \textsc{KuafuDet}, using real-time streaming, on a server with 16 GB memory, quad-core i7-4800MQ at 3.6 GHz, and 1 TB hard drives.

\subsection{Experimental Results}\label{experimentresults}
For a meaningful comparison, we list the results that are used to train on different classifiers with respect to the aspects of false negative (denoted as FN) and accuracy. FN rate refers to all malicious instances that are classified as benign applications. Accuracy simply measures that the classifier makes the correct prediction. Because we use our classifier as a tool for prioritizing the response to Android malware disclosures, we focus on improving the accuracy and reducing the false negative.

\subsubsection{Evaluation on attacks against the detection}\label{mamadroid}
Our collected dataset (16,000 samples as a training set and 4,000 samples as a test set, shown in Table~\ref{fig:dataset}) serves as a benchmark for evaluating robustness of Android malware detection systems. As mentioned, \textsc{DroidAPIMiner}, \textsc{Drebin}, and \textsc{MaMaDroid} are the three most recent Android malware detection systems. Since Support Vector Machine (SVM) is the only jointly-used algorithm by three detection systems, to conduct a fair comparison, we adopt SVM to investigate the misclassification rate of the three detection systems together with ours (when the adversarial detection mechanism is not included) under poisoning attacks. We first show that by poisoning their training set, it is possible to mislead their classifiers, along with ours; we then analyze the robustness of our discriminative classifiers against the three distinct attack strategies.
\vspace{2mm}

{\bf \noindent (\rmnum{1}) Misclassification of Machine Learning Detection Systems.} 

To perform a longitudinal study, we first apply our poisoning attack to \textsc{DroidAPIMiner}, \textsc{Drebin}, \textsc{MaMaDroid}, and \textsc{KuafuDet} without adversarial detection (Without AD). Specifically, \textit{Without AD} means that the camouflage detector component is disabled. We mimic sophisticated attacks to investigate how ineffective these systems perform under our poisoning attack. We assume to have control over a subset of samples and automatically generate the crafted camouflage samples as follows: (\rmnum{1}) we can only add or remove features. We must preserve the utility of the modified application, which we achieve by only adding features from benign set, and only those that do not interfere with the functionality of the application. (\rmnum{2}) We can add a restricted number of features. We thus validate that adversarial attack is indeed viable in security critical domains.

More specifically, we customize the adversarial crafting algorithm~\citep{papernot2016limitations} according to our adversarial model, to indicate the probability of adding or removing a specific feature. Therefore, for machine learning mechanisms that are based on syntax features, such as \textsc{DroidAPIMiner} and \textsc{Drebin}, we can directly apply our customized adversarial crafting algorithm; for machine learning mechanisms that mainly consider semantic and behavioral features, such as \textsc{MaMaDroid}, which relies on application behavior using Markov chain modeling, we instead target on crafting its feature in terms of call sequences. Specifically, we first extract a set of call sequences that are only frequently used by benign samples, and then add them to the the malicious samples, to mimic our sophisticated attack.

\begin{table}[t]\small
\caption{Misclassification rate comparison of adversarial detection}
\label{fig:c}
 \begin{center}
\begin{tabular}{|l|c|c|c|c|c|}
\hline
\multicolumn{2}{|l|}{\bf Detection Tool }& \textsc{\bf DroidAPIMiner} & \textsc{\bf Drebin}  & \textsc{\bf MaMaDroid}  & \textsc{\bf KuafuDet (Without AD)}\\ \hline
\hline
\multicolumn{2}{|l|}{\bf Misclassification rate (FN)} & 80.05\%  & 75.20\%  & 68.95\%  &62.60\%   \\ \hline
\end{tabular}
\end{center}
\end{table}

As shown in Table~\ref{fig:c}, we obtain 80.05\%, 75.20\%, and 68.95\% misclassification rates (\textit{i.e.}, FN) on \textsc{DroidAPIMiner}, \textsc{Drebin},  and \textsc{MaMaDroid}, respectively. We show that our poisoning attack on SVM is successfully validated through three machine-learning tools.

We here take a detailed discussion on the misclassification rate of the systems: 

\begin{itemize}
\item Our sophisticated attackers are able to mislead the machine learning detection systems.

\item \textsc{MaMaDroid} relies on transitional call sequences, rather than single API calls, to train its classifier, merely inserting syntax features as we did for poisoning other systems is not considered as a successful attack by our crafting algorithm. We thereby manipulate its feature space through call sequences.

\item \textsc{MaMaDroid} achieves lower misclassification rates than \textsc{Drebin} and \textsc{DroidAPIMiner} in the sophisticated attacks, sacrificing more computational time cost over each application due to call graph construction and feature extraction. Furthermore, \textsc{MaMaDroid} requires a sizable amount of memory when performing classification because of its large feature sets and the extraction of call graph.

\item \textsc{KuafuDet} (Without AD) also can be attacked through sophisticated attacks and the misclassification rate (62.60\%) indicates that it still suffers from adversarial samples.

\item \textsc{DroidAPIMiner}, \textsc{Drebin}, and \textsc{MaMaDroid} can be thwarted if we embed native code (as a strong attacker defined in our threat model) and dynamic code loading with reflection (as a sophisticated attack), because malicious code is loaded or determined at runtime. The attackers can pollute training data using a large-scale crafted samples through these techniques. 

\end{itemize}

In conclusion, the state-of-art machine learning-based malware detection systems are possible to be misled by the poisoning attacks we exhibit in the paper. By their nature, classifiers based on syntax features are more vulnerable than the ones based on semantic features. On the other hand, semantic features extraction does require more computational costs than syntax features. Hence, statistical robust features, pruned by information gain, are adopted in \textsc{KuafuDet} to accommodate scalable, generic, and large-scale malware detection.
In addition, \textsc{KuafuDet} provides specific mechanisms to countermeasure the dynamic loading and native code embedding poisonings. \textsc{KuafuDet} parses the native code and dynamic code from package, and then extracts the corresponding features to keep the robustness of classifiers.

In summary, we conjecture that almost all the state-of-the-art machine-learning malware detection systems are suffering from the poisoning attack we exhibited in the paper. 

\vspace{2mm}
{\bf \noindent (\rmnum{2}) Robustness of \textsc{KuafuDet}.}

Here, we analyze the robustness of our discriminative classifiers when encountering three distinct attack strategies. The first attack strategy is to launch a causative attack without any knowledge of the training data or ground truth. This \textit{weak attacker} in principle amounts to injecting noise into the system. The second attack strategy corresponds to the \textit{strong attacker}, who only manipulates partial features in the training set. The third, the most aggressive attacker we consider is the \textit{sophisticated attacker}. This attacker can fully manipulate almost all training features to launch a sophisticated attack, which creates scenarios where relatively benign mobile applications and real-world malicious mobile applications appear to have nearly identical attributes at the training stage. 

\begin{table*}[!htb] \renewcommand{\arraystretch}{1.2}
\scriptsize
\caption{The performance of adversarial detection}
\label{fig:result2}
 \begin{center}
\begin{tabular}{|l|c|c|c|c|c|c|c|c|c|}
\hline
{\bf Conventional  Detection} 		& \multicolumn{3}{c|}{SVM}      	& \multicolumn{3}{c|}{RF} 			& \multicolumn{3}{c|}{KNN} \\ \hline
FN           						& \multicolumn{3}{c|}{4.90\%}     	& \multicolumn{3}{c|}{2.50\%} 		& \multicolumn{3}{c|}{3.40\%} \\ \hline
Accuracy          					& \multicolumn{3}{c|}{94.95\%}     	& \multicolumn{3}{c|}{96.35\%} 		& \multicolumn{3}{c|}{95.80\%} \\ \hline

{\bf Attacker}       	 			& Weak & Strong & Sophisticated  & Weak & Strong& Sophisticated 		& Weak&Strong&Sophisticated \\ \hline
{\bf Without AD}            			& \multicolumn{3}{c|}{SVM} 		& \multicolumn{3}{c|}{RF} 			& \multicolumn{3}{c|}{KNN} \\ \hline
FN           						&8.60\%   &49.80\%&62.60\%    	&5.60\%&41.80\%&55.90\%          		&5.90\%  &26.20\% &45.40\% \\ \hline
Accuracy           					&93.10\% &72.50\%&\bf 64.30\%       	&94.80\%&76.40\%&\bf 67.85\%			&94.55\%&84.40\%&\bf 72.00\% \\ \hline
{\bf Within AD}             			& \multicolumn{3}{c|}{SVM} 		&\multicolumn{3}{c|}{RF} 			& \multicolumn{3}{c|}{KNN} \\ \hline
FN &5.80\% &10.00\%&12.60\%  		
&4.70\%&11.20\% &14.50\%  			
&4.10\%&9.20\%&11.60\% \\ \hline

Accuracy &94.50\% &92.40\%&\bf 89.30\%     	
&95.65\%&92.00\% &\bf 88.85\%		
&95.45\%&92.90&\bf 90.45\% \\ \hline
\end{tabular}
\end{center}
\end{table*}

\begin{figure*}[!htb]\small
\begin{minipage}[t]{0.5\linewidth}
\centering
\includegraphics[width=2.7in]{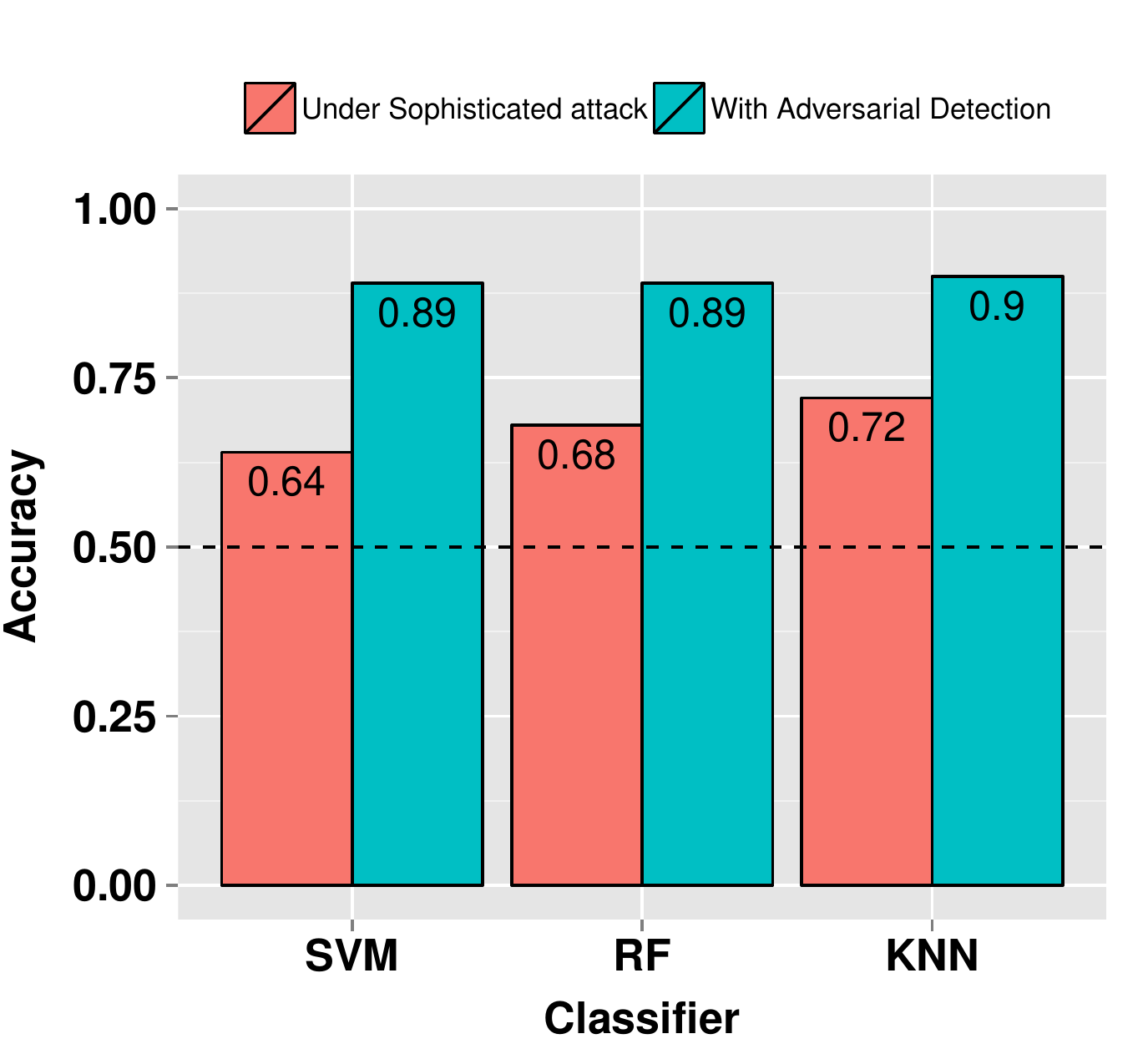}
\captionsetup{justification = centering, margin = 3mm}
\caption{Detection accuracy}
\label{fig:compare1}
\end{minipage}%
\begin{minipage}[t]{0.5\linewidth}
\centering
\includegraphics[width=2.7in]{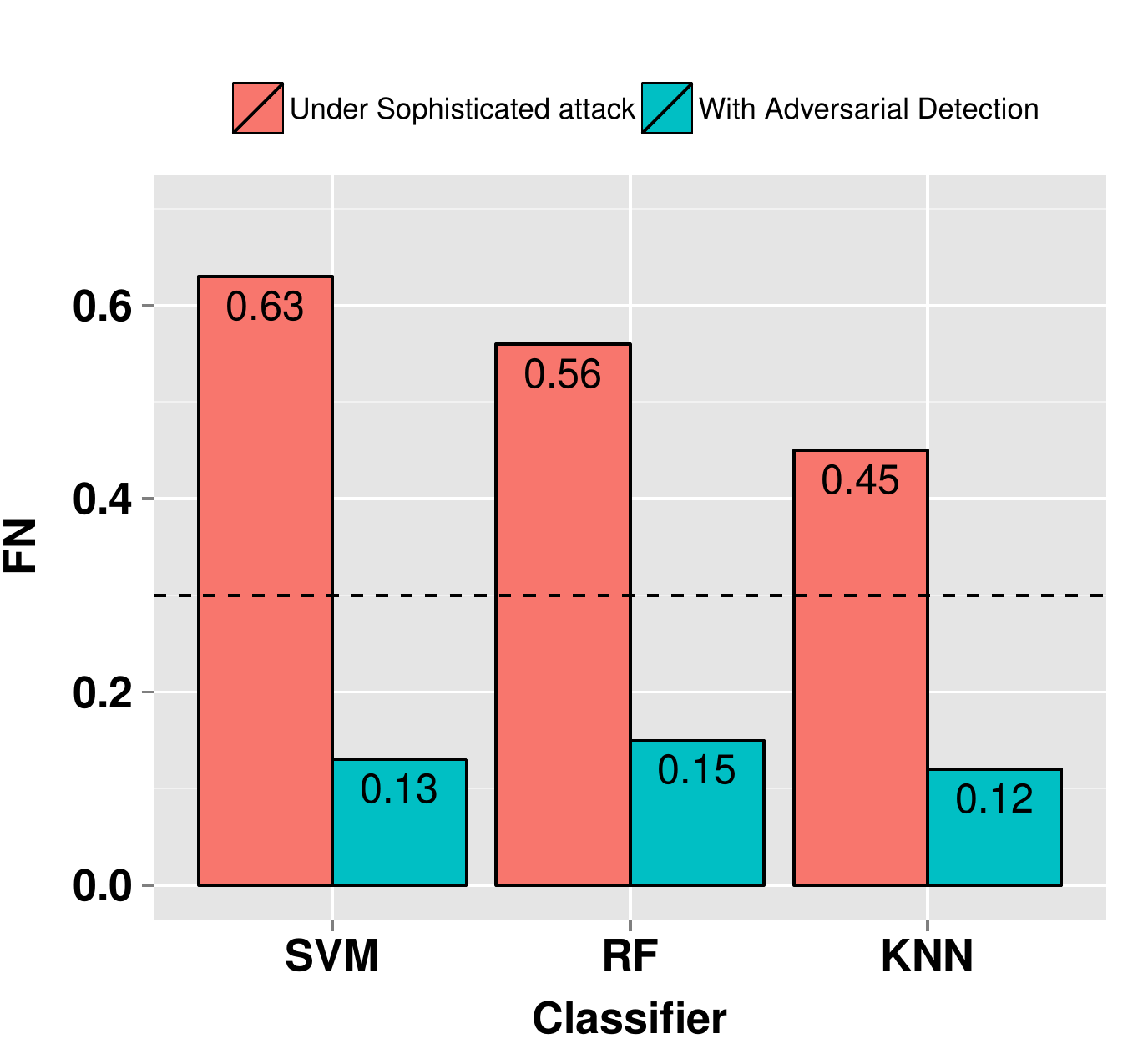}
\captionsetup{justification = centering, margin = 3mm}
\caption{Detection false negatives}
\label{fig:compare2}
\end{minipage}
\end{figure*}


\begin{figure*}[!htb]
  \centering
  \includegraphics[width=0.5\textwidth]{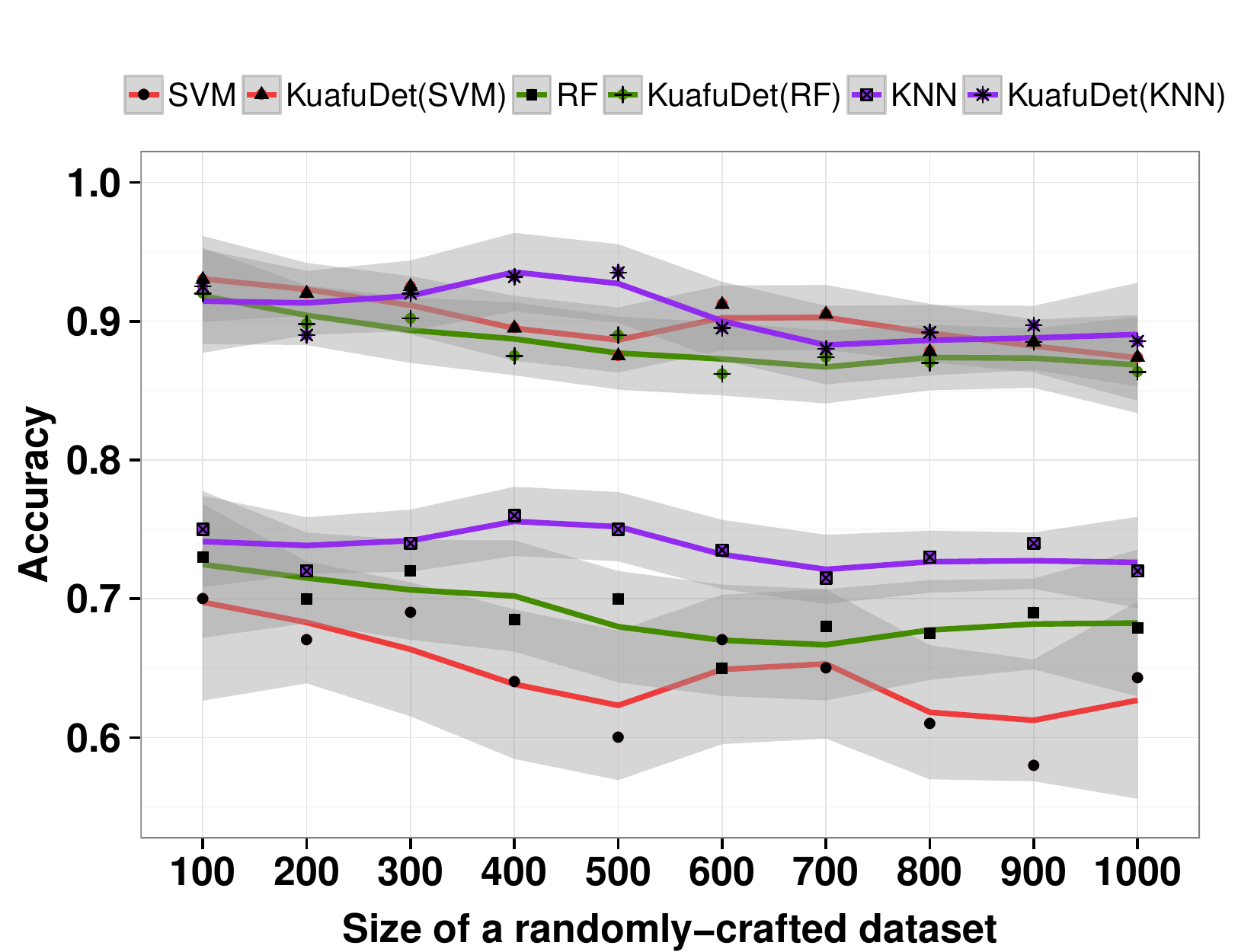}
  \caption{The robustness of \textsc{KuafuDet} (accuracy)}
  \label{r1}
\end{figure*}

As shown in Table~\ref{fig:result2}, the weak attacker is not able to force the accuracy of our malware detection below 90\%.
This suggests that discriminative classifiers can be relatively robust to this type of random noise-based attack. When dealing with the strong attacker
, performance degrades to approximately 90\% accuracy. The sophisticated attacker can cause the accuracy to drop to approximately 65\% by incorporating thousands of training set (green bar in Figure~\ref{fig:compare1}).
The sophisticated attacker represents a practical upper bound for the accuracy loss that a realistic attacker can inflict on our detection system. We see that injecting carefully crafted data into training data can significantly reduce detection accuracy. However, with the help of adversarial detection, holistic performance upgrades by at least 15\% accuracy with respect to each listed classifier. Hence, performance of our adversarial detection remains above baseline levels listed in~\citep{meng2016mystique} even for our strongest attackers due to the use of similarity-based filtering to increase classifier robustness. Analysis on false negatives has an analogous interpretation (see Figure~\ref{fig:compare2}). As shown in Table~\ref{fig:result2}, the algorithm KNN outperforms other algorithms under adversarial environment because of its higher resistance to random classification noise, which is aligned with the conclusion drawn from the recent research~\citep{wang2017analyzing}.


Figure~\ref{r1} displays scatter-plots of system accuracy ({\footnotesize $\left. (TP + TN) \middle/ (TP + TN + FP + FN) \right.$}) as a function of the size of randomly-crafted datasets, where TP and TN denote the number of samples correctly classified as malicious and benign, respectively. FP and FN indicate the number of samples mistakenly identified as malicious and benign, respectively. To ease presentation, the plots are fitted by Loess curves with $95\%$ confidence interval bands that depict the upper and lower confidence bounds for all points within the range of data, making it especially useful for comparing groups for which no theoretical models exist. As can be seen from Figure~\ref{r1}, as the dataset size grows, the accuracy drops slightly. This is somewhat expected as more potential evasion takes place. Furthermore, as discussed earlier, \textsc{KuafuDet} offers significantly better robustness of detection accuracy, regardless of either the size of the dataset or the type of classifier applied. Analysis on the robustness of recall has an analogous interpretation.

\subsubsection{Evaluation on Accuracy}
We here compare our work with the previous work with respect to accuracy. We opt to apply a large portion of our dataset used in \textsc{KuafuDet}, which intergrates the dataset owned by \textsc{Drebin}~\citep{arp2014drebin} and \textsc{StormDroid}~\citep{chen2016stormdroid}, in order to evaluate accuracy performance. Because the dataset used in \textsc{DroidAPIMiner} and \textsc{MaMaDroid} are not publicly available, to be fair, we do not apply the dataset that is not even used in \textsc{DroidAPIMiner} and \textsc{MaMaDroid} \textit{per se} to them to have an asymmetric advantage over it. Hence, \textsc{Drebin} and \textsc{StormDroid} are only considered. As shown in Table~\ref{fig:result1}, our accuracy rate (96.35\%) completely outperforms the accuracy rate in \textsc{StormDroid} (93.80\%)~\citep{chen2016stormdroid} and \textsc{Drebin} (93.90\%)~\citep{arp2014drebin}, let alone use our combined dataset.

As evidenced by Table~\ref{fig:result1}, we achieve the highest accuracy because of the feature selection and similarity-based filtering.

\begin{table}[!htb]\footnotesize
\caption{Accuracy comparison}
\label{fig:result1}
\begin{minipage}{\columnwidth}
 \begin{center}
 \begin{tabular}{|l|c|c|}
 \hline {\bf  Detection Tool}  & {\bf Accuracy} & {\bf \# Malware}\\
 \hline
 \hline \textsc{KuafuDet} 	&  {\bf 96.35\%} & 10,400 \\
 \hline \textsc{Drebin} 		& 93.90\%  & 5,560 \\
 \hline \textsc{StormDroid} 	& 93.80\%  & 3,620 \\
 \hline
\end{tabular}
\end{center}
\end{minipage}
\end{table}

\subsubsection{Robustness of Imbalanced Data}
Our experiments aim to evaluate the robustness of \textsc{KuafuDet} when the ratio goes imbalanced in the adversarial environment. To evaluate the robustness of \textsc{KuafuDet} in the data-imbalanced environment, we first apply 4,000 malicious and 4,000 benign apps (\textit{i.e.}, 1:1 ratio) for training our classifiers, and gradually add benign ones to achieve different ratios up to 1:50. To be specific, we conduct experiments with ratios including 1:1, 1:5, 1:10, 1:20, and 1:50. We use 10-fold cross-validation in our experiments. As shown in Table~\ref{fig:ratios}, the accuracy degrades as we approach real-world ratio of malware and benign apps, but the accuracy still remains above 90\%. In the self-adaptive scheme, we, nevertheless, show the capability to tackle the problem of the imbalanced data ratio.

\begin{table}[!htb]\small
\caption{Results for different malware to benign apps ratios}
\label{fig:ratios}
 \begin{center}
\begin{tabular}{|l|c|c|c|c|c|c|}
\hline
\multicolumn{2}{|l|}{\bf Ratio}& {\bf 1:1} & {\bf 1:5} & {\bf 1:10} & {\bf 1:20} & {\bf 1:50} \\ \hline
\hline
\multicolumn{2}{|l|}{\bf Accuracy} &96.40\%   & 96.15\%  &95.80\%   &94.60\% &93.75\%       \\ \hline
\end{tabular}
\end{center}
\end{table}

\subsubsection{Evaluation on Time Cost, Scalability, and Adaptability}
To support a high-performance malware detection, \textsc{KuafuDet} is designed to run on top of an open-source distributed stream-processing engine \textit{Storm}.\footnote{\url{http://storm.apache.org/}} \textsc{KuafuDet} supports a large-scale detection of a data stream by a set of worker units that connect to each other, forming a topology: A submitted application is first disassembled to extract its features; then, the metrics-driven pruning and information gain analysis are run, and two-phase iterative adversarial detection is finally activated. Each operation here is delegated to a worker unit on the topology and all the data associated with the application are in a single stream. Running on top of the \textit{Storm} stream processor, \textsc{KuafuDet} is tested on the platform of Pwnzen Infotech Inc. We show that the size of the test group is unaffected with the efficiency of our tool.  As shown in Table~\ref{fig:time1}, average detection time per application is less than 3 seconds, which is indeed capable of scaling up to the massive datasets.

\begin{table}[!htb]\footnotesize
\caption{Time cost of \textsc{KuafuDet} in units of seconds}
    \label{fig:time1}
   \begin{minipage}{\columnwidth}
 \begin{center}
 \begin{tabular}{|c|c|c|}
 \hline {\bf \# APKs}&{\bf Total Time} &{\bf AVG Time/APK }\\
 \hline
 \hline  200 	& 518 	& 2.59 \\
 \hline  400 	& 1066 	& 2.67 \\
 \hline  600  	& 1578 	& 2.63 \\
 \hline  800  	& 2097 	& 2.62 \\
 \hline  1,000 	& 2778 	& 2.78 \\
 \hline
\end{tabular}
\end{center}
\end{minipage}
\end{table}

\subsubsection{Evaluation on Coverage}
To circumvent the over-fitting issue and to better understand the coverage of \textsc{KuafuDet}, we randomly sample 1,000 malicious applications from 217 Android malware families from our dataset to cover almost all the existing Android malicious behaviors. We scan them using \textsc{KuafuDet} and other well-known industrial malware detection tools, such as Kaspersky and McAfee encapsulated in VirusTotal. Although the coverage of \textsc{KuafuDet}, with the combined top features, is 96.20\%, better than what can be achieved by any individual scanner, including such top-of-the-line antivirus systems as ESET-NOD32 (79.50\%), McAfee (75.50\%), Ikarus (72.50\%), Kaspersky (72.10\%), and Avira (69.30\%), industrial tools deal with millions of applications, many of which are zero-day. We argue that comparing with industrial tools is to understand the different emphases in academia and industry. The breakdowns of the coverage study is presented in Table~\ref{fig:tools}.

\begin{table}[!htb]\scriptsize
\caption{Coverage comparison}
\label{fig:tools}
\begin{minipage}{\columnwidth}
 \begin{center}
 \begin{tabular}{|l|c|}
 \hline {\bf  Detection Tool}  & {\bf Percentage}  \\
 \hline
 \hline \textsc{KuafuDet} 	  	& {\bf 96.20\%} \\
  \hline ESET-NOD32  	   		& 79.50\%  \\
  \hline McAfee  				& 75.50\% \\
  \hline Ikarus 			  	& 72.50\%  \\
  \hline Kaspersky  		         & 72.10\% \\
  \hline Avira  				& 69.30\% \\
  \hline VIPER  				& 67.50\%  \\
  \hline Qihoo-360  		  	& 62.30\% \\
  \hline Symantec  		   	& 40.40\%  \\
  \hline
\end{tabular}
\end{center}
\end{minipage}
\end{table}

Since \textsc{KuafuDet} decouples the similarity-based filtering from machine learning classifiers, it enables us to periodically enhance the learning system. Moreover, \textsc{KuafuDet} also considers an attacker threat dimension, which makes the whole system design completely adaptable to new malware.

\section{Discussion}\label{discussion}
Our study is limited in five ways as discussed in the following.

{\bf \noindent (\rmnum{1}) The granularity of classifiers.}
The hyperplane between benign and malicious can be blurred and subjective, which depends on specific security requirements and uses cases to determine whether a pattern is really benign or malicious. For example, if individual users root their own devices and use the game hacking applications, game developers is very likely to treat them as malicious because they bypass the in-app purchase. In practice, this kind of apps is defined as ``grayware'' that has no clear distinctive difference between the benign and the malicious. ``Grayware'' is now becoming a great threat to mobile devices since attackers achieve more profit in this way. \textsc{KuafuDet} is a generic and coarse-grained architecture for classifying applications with high accuracy. As for such grayware, \textsc{KuafuDet} can be tuned for a specific detection.

{\bf \noindent (\rmnum{2}) The limitation of decompilation technologies.}
We extract features from the \textit{manifest} and \textit{Smali} files that are successfully decompiled. However, in our experiments, we find that a few APK files cannot be decompiled successfully. As for this situation, we change the decompiling tool in our experiments. For future study, we will explore the possibility to use reinforcement techniques to prevent the APKs from reverse-engineering. This would increase the difficulty of unpacking the original APK for attackers.

{\bf \noindent (\rmnum{3}) The scarcity of empirical samples.}
Although we find some case studies reflecting the different attacker levels for machine learning classifiers, we still lack a huge number of samples to scrutinize. We note attackers might be able to deliberately force the benign and malicious access patterns to co-occur in one log, such as triggering the benign pattern first and then launching the attack, though we have not observed in the wild. This perhaps requires to dilute the poisoned logs and possibly requires human analysts to contribute external knowledge. We hope in future study that using Game Theory is beneficial to our interpretation of the attackers psychology so as to take targeted strategy to detect stubborn malware.

{\bf \noindent (\rmnum{4}) The limitation of selected features.} Although the 195 features are representative and extracted by using statistical metrics-driven manual pruning with information gain in our experiments, new malicious behaviors might disturb the feature space, making the system less effective. 

{\bf \noindent (\rmnum{5}) The limitation of similarity-based approach.} (1) Attackers can somehow approximate our similarity-based approach by inferring the similarity thresholds used. However, it is actually difficult to infer our selected samples that are used to calculate the thresholds. Furthermore, the thresholds will change as selected samples update over time, for which attackers will take great efforts to exploit. (2) \textsc{KuafuDet}, through a self-adaptive learning scheme, discovers new information from both the identified malware and the filtered suspicious false negatives from camouflage detector. We acknowledge that this process would cause false positives. For example, SMS-related applications are benign applications, but they also have sensitive behaviors that are typical in Android malware.

\section{Related Work}\label{related}
Contemporary machine learning-based techniques typically model the detection problem as a binary classification problem. Together with system analysis techniques, the malicious behaviors can be studied and employed to increase their detection performance, especially for mobile applications in the wild.

\subsection{Machine Learning-based Detection}
Arp \textit{et al.}~\citep{arp2014drebin} built the \textsc{Drebin} system, which works with a massive feature set extracted from the manifest file and the app's source code and trains an SVM classifier for malware detection. Although \textsc{Drebin} has accommodated thousands of features with an impressive performance results, it suffers two challenges: first, the malware is out-dated and well recorded in malware detection tools; second, a comprehensive coverage of different attacking and evasion techniques is missing. 

\textsc{DroidAPIMiner}~\citep{aafer2013droidapiminer} mainly extracts the top 169 API calls, which are used more frequently in the malware than in the benign set, package level information, as well as some dangerous parameter information as features to analyze a large corpus of Android malware. Because of the evolution of both malware and the Android API, it requires constant retraining on most common calls. 

Most recently, Chen \emph{et al.}~\cite{chen2016more} suggest the use of semantic features of mobile apps to retain classifier value over time, building on the intuition that certain semantic attributes of mobile malware are invariant. Experiments verify that the incorporation of semantic features can significantly improve the performance of Android malware classification. Deo~\emph{et al.}~\cite{deo2016prescience} propose to assess the quality of binary classification by using probabilistic predictors. Although they both consider retraining, adversarial environment is missing. 

MUDFLOW~\citep{avdiienko2015mining} argues that the pattern of sensitive information flows in malware is statistically different from those in benign apps. From an application, MUDFLOW uses static analysis to extract the flow paths, and these flow paths are then mapped to a feature vector used in a specific classifier. DroidSIFT~\citep{zhang2014semantics} is unique in designing features in terms of the generation of API dependency graphs $G$ for each app, and the construction of the feature vector of the app. The features represent the similarity of the graphs $G$ corresponding to the database of graphs of known benign apps and malware apps. Finally, the feature vectors are used in anomaly detection. However, the dataset for detection is not large enough, yielding low effectiveness and confidence in classification. 
DroidMat~\citep{wu2012droidmat} uses a static feature-based mechanism, which considers static information (\textit{e.g.}, permission, intent messages, API calls) for detecting Android malware. It uses $K$-means algorithm to enhance the malware modeling capability and KNN algorithm to classify apps as benign or malicious. However, it does not extract semantic features for training and does not take adversarial environment into consideration. 
Shabtai \textit{et al.}~\citep{shabtai2012andromaly} presented a host-based malware detection system that continuously monitors mobile devices to detect malicious data using a supervised machine learning anomaly detection technique. It focuses on host-based malware, while our approach focuses on mobile malware detection in adversarial environment. 
Most recently, \textsc{MaMaDroiD}~\citep{mariconti2016mamadroid} is built to maintain resilience to API changes, but it requires a large amount of memory when performing classification and a substantial amount of time per app.

\subsection{Evasion Techniques}
Currently, the issues of understanding machine learning security in adversarial settings mainly focus on spam email detection~\citep{zhang2016adversarial, bruckner2012static, biggio2014security, debarr2013adversarial, wang2014sparse}. Recently, many statistical adversarial models are proposed to construct effective adversarial samples, such as using deep neutral networks~\citep{papernot2016limitations, grosse2016adversarial, mcdaniel2016machine, li2016adversarial, shen2016uror}. As seen from a generic perspective, Wang \textit{et al.}~\citep{wang2016theoretical} utilized the notation of topological spaces and oracles to explain why an adversarial sample can bypass a classifier, and they generated a sufficient and necessary condition to determine the robustness of classifiers under adversarial environment. Goodfellow \textit{et al.}~\citep{goodfellow2014explaining} explained and generated adversarial samples for adversarial training to reduce test error. 
However, all of these studies did not focus on specific causation leading to evasion in the mobile malware context. They also did not show the feasibility how these adversarial samples work in the wild.

For conventional malware evasion, one straightforward evasion technique is to repackage a benign app with small snippets of malicious code added to several classes. Moreover, attackers could also use reflection, dynamic code loading, or native code~\citep{poeplau2014execute}. Such attempts to escape detection are likely to be deemed suspicious. Among them, DroidChameleon~\citep{rastogi2014catch} integrates three types of transformation techniques and generates obfuscated mobile malware. Mystique~\citep{meng2016mystique} develops a framework to automatically generate malware covering four attack features and two evasion features to obfuscate the generated malware. For the general defense, Cao~\textit{et al.}~\citep{cao2015towards} presented a proof-of-concept machine unlearning prototype that can rapidly forget data to regain privacy, security, and usability. The current paper is an extension of a poster~\citep{ chen2016towards}.

In summary, previous work either conducts evasion techniques without considering the feature space or only using machine learning-based approaches. With our experiments and real-world case studies, it is obvious that attacks can also poison features while preserving maliciousness, and our experiments verified that the resulting fake variants with poisoned features impaired discriminative classifiers and succeeded in lowering the detection score in a test environment. To the best of our knowledge, this is one of the first papers to accommodate adversarial machine learning into mobile malware detection. We are also the first paper to show the possibility to defend against adversarial attacks on mobile malware, to the greatest extent.

\section{Conclusion}\label{conclusion}
We reviewed several challenges for the malware detection problem. We showed how the conventional machine learning classifiers can fail against determined attackers. Based on these insights, we designed and evaluated three types of attackers targeting the training phases to poison our detection. Through simulation, we presented practical bounds for the accuracy loss to each attacker. To address this threat, we therefore proposed our detection system, \textsc{KuafuDet}, and showed it significantly reduces false negatives and boosts the detection accuracy by at least 15\%. 

We argue that it is essential to inform researchers considering how attackers will adapt to the conventional detection, as well as to inform developers working on the next-generation malware detection systems. We conjecture that the arms race will be over only when the effectiveness of early detection will sufficiently increase the cost of infection.

\bibliography{sample}

\begin{thebibliography}{54}
\expandafter\ifx\csname natexlab\endcsname\relax\def\natexlab#1{#1}\fi
\expandafter\ifx\csname url\endcsname\relax
  \def\url#1{\texttt{#1}}\fi
\expandafter\ifx\csname urlprefix\endcsname\relax\def\urlprefix{URL }\fi

\bibitem[{Aafer et~al.(2013)Aafer, Du, and Yin}]{aafer2013droidapiminer}
Aafer, Y., Du, W., Yin, H., 2013. {DroidAPIMiner}: Mining {API}-level features
  for robust malware detection in {Android}. In: Security and Privacy in
  Communication Networks. Springer, pp. 86--103.

\bibitem[{Arp et~al.(2014)Arp, Spreitzenbarth, Hubner, Gascon, and
  Rieck}]{arp2014drebin}
Arp, D., Spreitzenbarth, M., Hubner, M., Gascon, H., Rieck, K., 2014. {DREBIN}:
  Effective and explainable detection of {Android} malware in your pocket. In:
  Proceedings of the Annual Symposium on Network and Distributed System
  Security (NDSS).

\bibitem[{Arzt et~al.(2014)Arzt, Rasthofer, Fritz, Bodden, Bartel, Klein,
  Le~Traon, Octeau, and McDaniel}]{arzt2014flowdroid}
Arzt, S., Rasthofer, S., Fritz, C., Bodden, E., Bartel, A., Klein, J.,
  Le~Traon, Y., Octeau, D., McDaniel, P., 2014. Flowdroid: Precise context,
  flow, field, object-sensitive and lifecycle-aware taint analysis for
  {Android} apps. In: ACM SIGPLAN Notices. Vol.~49. ACM, pp. 259--269.

\bibitem[{Avdiienko et~al.(2015)Avdiienko, Kuznetsov, Gorla, Zeller, Arzt,
  Rasthofer, and Bodden}]{avdiienko2015mining}
Avdiienko, V., Kuznetsov, K., Gorla, A., Zeller, A., Arzt, S., Rasthofer, S.,
  Bodden, E., 2015. Mining {Apps} for abnormal usage of sensitive data. In:
  2015 IEEE/ACM 37th IEEE International Conference on Software Engineering.
  Vol.~1. IEEE, pp. 426--436.

\bibitem[{Biggio et~al.(2014)Biggio, Fumera, and Roli}]{biggio2014security}
Biggio, B., Fumera, G., Roli, F., 2014. Security evaluation of pattern
  classifiers under attack. IEEE Transactions on Knowledge and Data Engineering
  26~(4), 984--996.

\bibitem[{Br{\"u}ckner et~al.(2012)Br{\"u}ckner, Kanzow, and
  Scheffer}]{bruckner2012static}
Br{\"u}ckner, M., Kanzow, C., Scheffer, T., 2012. Static prediction games for
  adversarial learning problems. Journal of Machine Learning Research 13~(Sep),
  2617--2654.

\bibitem[{Brunk(2009)}]{brunk2009intersection}
Brunk, F., 2009. Intersection problems in combinatorics. Ph.D. thesis,
  University of St Andrews.

\bibitem[{Cao and Yang(2015)}]{cao2015towards}
Cao, Y., Yang, J., 2015. Towards making systems forget with machine unlearning.
  In: Security and Privacy (SP), 2015 IEEE Symposium on. IEEE, pp. 463--480.

\bibitem[{Chen et~al.(2016{\natexlab{a}})Chen, Xue, Tang, Xu, and
  Zhu}]{chen2016stormdroid}
Chen, S., Xue, M., Tang, Z., Xu, L., Zhu, H., 2016{\natexlab{a}}. Stormdroid: A
  streaminglized machine learning-based system for detecting {Android} malware.
  In: Proceedings of the 11th ACM on Asia Conference on Computer and
  Communications Security. ACM, pp. 377--388.

\bibitem[{Chen et~al.(2016{\natexlab{b}})Chen, Xue, and Xu}]{chen2016towards}
Chen, S., Xue, M., Xu, L., 2016{\natexlab{b}}. Poster: Towards adversarial
  detection of mobile malware. In: Proceedings of the 22nd Annual International
  Conference on Mobile Computing and Networking. ACM, pp. 415--416.

\bibitem[{Chen et~al.(2016{\natexlab{c}})Chen, Aspinall, Gordon, Sutton, and
  Muttik}]{chen2016more}
Chen, W., Aspinall, D., Gordon, A.~D., Sutton, C., Muttik, I.,
  2016{\natexlab{c}}. More semantics more robust: Improving android malware
  classifiers. In: Proceedings of the 9th ACM Conference on Security \& Privacy
  in Wireless and Mobile Networks. ACM, pp. 147--158.

\bibitem[{Dash et~al.(2016)Dash, Suarez-Tangil, Khan, Tam, Ahmadi, Kinder, and
  Cavallaro}]{dash2016droidscribe}
Dash, S.~K., Suarez-Tangil, G., Khan, S., Tam, K., Ahmadi, M., Kinder, J.,
  Cavallaro, L., 2016. {DroidScribe}: Classifying {Android} malware based on
  runtime behavior. Mobile Security Technologies (MoST 2016) 7148, 1--12.

\bibitem[{Debarr et~al.(2013)Debarr, Sun, and Wechsler}]{debarr2013adversarial}
Debarr, D., Sun, H., Wechsler, H., 2013. Adversarial spam detection using the
  randomized hough transform-support vector machine. In: Machine Learning and
  Applications (ICMLA), 2013 12th International Conference on. Vol.~1. IEEE,
  pp. 299--304.

\bibitem[{Deo et~al.(2016)Deo, Dash, Suarez-Tangil, Vovk, and
  Cavallaro}]{deo2016prescience}
Deo, A., Dash, S.~K., Suarez-Tangil, G., Vovk, V., Cavallaro, L., 2016.
  Prescience: Probabilistic guidance on the retraining conundrum for malware
  detection. In: Proceedings of the 2016 ACM Workshop on Artificial
  Intelligence and Security. ACM, pp. 71--82.

\bibitem[{Enck et~al.(2014)Enck, Gilbert, Han, Tendulkar, Chun, Cox, Jung,
  McDaniel, and Sheth}]{enck2014taintdroid}
Enck, W., Gilbert, P., Han, S., Tendulkar, V., Chun, B.-G., Cox, L.~P., Jung,
  J., McDaniel, P., Sheth, A.~N., 2014. {TaintDroid}: an information-flow
  tracking system for realtime privacy monitoring on smartphones. ACM
  Transactions on Computer Systems (TOCS) 32~(2), 5.

\bibitem[{Fan et~al.(2016)Fan, Xue, Chen, Xu, and Zhu}]{fan2016poster}
Fan, L., Xue, M., Chen, S., Xu, L., Zhu, H., 2016. Poster: Accuracy vs. time
  cost: Detecting android malware through pareto ensemble pruning. In:
  Proceedings of the 2016 ACM SIGSAC Conference on Computer and Communications
  Security. ACM, pp. 1748--1750.

\bibitem[{Feizollah et~al.(2017)Feizollah, Anuar, Salleh, Suarez-Tangil, and
  Furnell}]{feizollah2017androdialysis}
Feizollah, A., Anuar, N.~B., Salleh, R., Suarez-Tangil, G., Furnell, S., 2017.
  Androdialysis: Analysis of android intent effectiveness in malware detection.
  Computers \& Security 65, 121--134.

\bibitem[{Felt et~al.(2011)Felt, Chin, Hanna, Song, and
  Wagner}]{felt2011android}
Felt, A.~P., Chin, E., Hanna, S., Song, D., Wagner, D., 2011. Android
  permissions demystified. In: Proceedings of the 18th ACM conference on
  Computer and communications security. ACM, pp. 627--638.

\bibitem[{Goodfellow et~al.(2014)Goodfellow, Shlens, and
  Szegedy}]{goodfellow2014explaining}
Goodfellow, I.~J., Shlens, J., Szegedy, C., 2014. Explaining and harnessing
  adversarial examples. arXiv preprint arXiv:1412.6572.

\bibitem[{Gordon et~al.(2015)Gordon, Kim, Perkins, Gilham, Nguyen, and
  Rinard}]{gordon2015information}
Gordon, M.~I., Kim, D., Perkins, J.~H., Gilham, L., Nguyen, N., Rinard, M.~C.,
  2015. Information flow analysis of {Android} applications in {DroidSafe}. In:
  Proceedings of the Annual Symposium on Network and Distributed System
  Security (NDSS).

\bibitem[{Graziano et~al.(2015)Graziano, Canali, Bilge, Lanzi, and
  Balzarotti}]{graziano2015needles}
Graziano, M., Canali, D., Bilge, L., Lanzi, A., Balzarotti, D., 2015. Needles
  in a haystack: mining information from public dynamic analysis sandboxes for
  malware intelligence. In: 24th USENIX Security Symposium (USENIX Security
  15). pp. 1057--1072.

\bibitem[{Grosse et~al.(2016)Grosse, Papernot, Manoharan, Backes, and
  McDaniel}]{grosse2016adversarial}
Grosse, K., Papernot, N., Manoharan, P., Backes, M., McDaniel, P., 2016.
  Adversarial perturbations against deep neural networks for malware
  classification. arXiv preprint arXiv:1606.04435.

\bibitem[{Hall et~al.(2009)Hall, Frank, Holmes, Pfahringer, Reutemann, and
  Witten}]{hall2009weka}
Hall, M., Frank, E., Holmes, G., Pfahringer, B., Reutemann, P., Witten, I.~H.,
  2009. The {WEKA} data mining software: an update. ACM SIGKDD explorations
  newsletter 11, 10--18.

\bibitem[{Idrees et~al.(2017)Idrees, Rajarajan, Conti, Chen, and
  Rahulamathavan}]{idrees2017pindroid}
Idrees, F., Rajarajan, M., Conti, M., Chen, T., Rahulamathavan, Y., 2017.
  Pindroid: a novel android malware detection system using ensemble learning
  methods. Computers \& Security.

\bibitem[{Li et~al.(2015)Li, Bartel, Bissyand{\'e}, Klein, Le~Traon, Arzt,
  Rasthofer, Bodden, Octeau, and McDaniel}]{li2015iccta}
Li, L., Bartel, A., Bissyand{\'e}, T.~F., Klein, J., Le~Traon, Y., Arzt, S.,
  Rasthofer, S., Bodden, E., Octeau, D., McDaniel, P., 2015. {IccTA}: Detecting
  inter-component privacy leaks in {Android} apps. In: Proceedings of the 37th
  International Conference on Software Engineering. Vol.~1. IEEE Press, pp.
  280--291.

\bibitem[{Li and Li(2016)}]{li2016adversarial}
Li, X., Li, F., 2016. Adversarial examples detection in deep networks with
  convolutional filter statistics. arXiv preprint arXiv:1612.07767.

\bibitem[{Mariconti et~al.(2017)Mariconti, Onwuzurike, Andriotis,
  De~Cristofaro, Ross, and Stringhini}]{mariconti2016mamadroid}
Mariconti, E., Onwuzurike, L., Andriotis, P., De~Cristofaro, E., Ross, G.,
  Stringhini, G., 2017. {MAMADROID}: Detecting {Android} malware by building
  {Markov} chains of behavioral models. In: Proceedings of the Annual Symposium
  on Network and Distributed System Security (NDSS).

\bibitem[{McDaniel et~al.(2016)McDaniel, Papernot, and
  Celik}]{mcdaniel2016machine}
McDaniel, P., Papernot, N., Celik, Z.~B., 2016. Machine learning in adversarial
  settings. IEEE Security \& Privacy 14~(3), 68--72.

\bibitem[{Meng et~al.(2016)Meng, Xue, Mahinthan, Narayanan, Liu, Zhang, and
  Chen}]{meng2016mystique}
Meng, G., Xue, Y., Mahinthan, C., Narayanan, A., Liu, Y., Zhang, J., Chen, T.,
  2016. Mystique: Evolving {Android} malware for auditing anti-malware tools.
  In: Proceedings of the 11th ACM on Asia Conference on Computer and
  Communications Security. ACM, pp. 365--376.

\bibitem[{Octeau et~al.(2016)Octeau, Jha, Dering, McDaniel, Bartel, Li, Klein,
  and Le~Traon}]{octeau2016combining}
Octeau, D., Jha, S., Dering, M., McDaniel, P., Bartel, A., Li, L., Klein, J.,
  Le~Traon, Y., 2016. Combining static analysis with probabilistic models to
  enable market-scale android inter-component analysis. In: ACM SIGPLAN
  Notices. Vol.~51. ACM, pp. 469--484.

\bibitem[{Papernot et~al.(2016)Papernot, McDaniel, Jha, Fredrikson, Celik, and
  Swami}]{papernot2016limitations}
Papernot, N., McDaniel, P., Jha, S., Fredrikson, M., Celik, Z.~B., Swami, A.,
  2016. The limitations of deep learning in adversarial settings. In: Security
  and Privacy (EuroS\&P), 2016 IEEE European Symposium on. IEEE, pp. 372--387.

\bibitem[{Poeplau et~al.(2014)Poeplau, Fratantonio, Bianchi, Kruegel, and
  Vigna}]{poeplau2014execute}
Poeplau, S., Fratantonio, Y., Bianchi, A., Kruegel, C., Vigna, G., 2014.
  Execute this! analyzing unsafe and malicious dynamic code loading in android
  applications. In: NDSS. Vol.~14. pp. 23--26.

\bibitem[{Rasthofer et~al.(2014)Rasthofer, Arzt, and
  Bodden}]{rasthofer2014machine}
Rasthofer, S., Arzt, S., Bodden, E., 2014. A machine-learning approach for
  classifying and categorizing {Android} sources and {Sinks}. In: Proceedings
  of the Annual Symposium on Network and Distributed System Security (NDSS).

\bibitem[{Rasthofer et~al.(2016)Rasthofer, Arzt, Miltenberger, and
  Bodden}]{rasthofer2016harvesting}
Rasthofer, S., Arzt, S., Miltenberger, M., Bodden, E., 2016. Harvesting runtime
  values in {Android} applications that feature anti-analysis techniques. In:
  Proceedings of the Annual Symposium on Network and Distributed System
  Security (NDSS).

\bibitem[{Rastogi et~al.(2014)Rastogi, Chen, and Jiang}]{rastogi2014catch}
Rastogi, V., Chen, Y., Jiang, X., 2014. Catch me if you can: Evaluating
  {Android} anti-malware against transformation attacks. IEEE Transactions on
  Information Forensics and Security 9~(1), 99--108.

\bibitem[{Schlegel et~al.(2011)Schlegel, Zhang, Zhou, Intwala, Kapadia, and
  Wang}]{schlegel2011soundcomber}
Schlegel, R., Zhang, K., Zhou, X.-y., Intwala, M., Kapadia, A., Wang, X., 2011.
  Soundcomber: A stealthy and context-aware sound {Trojan} for smartphones. In:
  Proceedings of the Annual Symposium on Network and Distributed System
  Security (NDSS). Vol.~11. pp. 17--33.

\bibitem[{Shabtai et~al.(2012)Shabtai, Kanonov, Elovici, Glezer, and
  Weiss}]{shabtai2012andromaly}
Shabtai, A., Kanonov, U., Elovici, Y., Glezer, C., Weiss, Y., 2012. Andromaly:
  a behavioral malware detection framework for {Android} devices. Journal of
  Intelligent Information Systems 38~(1), 161--190.

\bibitem[{Shen et~al.(2016)Shen, Tople, and Saxena}]{shen2016uror}
Shen, S., Tople, S., Saxena, P., 2016. A uror: defending against poisoning
  attacks in collaborative deep learning systems. In: Proceedings of the 32nd
  Annual Conference on Computer Security Applications. ACM, pp. 508--519.

\bibitem[{Szegedy et~al.(2014)Szegedy, Zaremba, Sutskever, Bruna, Erhan,
  Goodfellow, and Fergus}]{szegedy2013intriguing}
Szegedy, C., Zaremba, W., Sutskever, I., Bruna, J., Erhan, D., Goodfellow, I.,
  Fergus, R., 2014. Intriguing properties of neural networks. In: Proceedings
  of the 2014 International Conference on Learning Representations.
  Computational and Biological Learning Society.

\bibitem[{Tam et~al.(2015)Tam, Khan, Fattori, and
  Cavallaro}]{tam2015copperdroid}
Tam, K., Khan, S.~J., Fattori, A., Cavallaro, L., 2015. {CopperDroid}:
  Automatic reconstruction of {Android} malware behaviors. In: Proceedings of
  the Annual Symposium on Network and Distributed System Security (NDSS).

\bibitem[{Wang et~al.(2016)Wang, Gao, and Qi}]{wang2016theoretical}
Wang, B., Gao, J., Qi, Y., 2016. A theoretical framework for robustness of
  (deep) classifiers under adversarial noise. arXiv preprint arXiv:1612.00334.

\bibitem[{Wang et~al.(2014)Wang, Liu, and Chawla}]{wang2014sparse}
Wang, F., Liu, W., Chawla, S., 2014. On sparse feature attacks in adversarial
  learning. In: Data Mining (ICDM), 2014 IEEE International Conference on.
  IEEE, pp. 1013--1018.

\bibitem[{Wang et~al.(2017)Wang, Jha, and Chaudhuri}]{wang2017analyzing}
Wang, Y., Jha, S., Chaudhuri, K., 2017. Analyzing the robustness of nearest
  neighbors to adversarial examples. arXiv preprint arXiv:1706.03922.

\bibitem[{Wong and Lie(2016)}]{wong2016intellidroid}
Wong, M.~Y., Lie, D., 2016. {IntelliDroid}: A targeted input generator for the
  dynamic analysis of android malware. In: Proceedings of the Annual Symposium
  on Network and Distributed System Security (NDSS).

\bibitem[{Wu et~al.(2014)Wu, Zhou, Patel, Liang, and Jiang}]{wu2014airbag}
Wu, C., Zhou, Y., Patel, K., Liang, Z., Jiang, X., 2014. {AirBag}: Boosting
  smartphone resistance to malware infection. In: Proceedings of the Annual
  Symposium on Network and Distributed System Security (NDSS).

\bibitem[{Wu et~al.(2012)Wu, Mao, Wei, Lee, and Wu}]{wu2012droidmat}
Wu, D.-J., Mao, C.-H., Wei, T.-E., Lee, H.-M., Wu, K.-P., 2012. Droidmat:
  {Android} malware detection through manifest and {API} calls tracing. In:
  Information Security (Asia JCIS), 2012 Seventh Asia Joint Conference on.
  IEEE, pp. 62--69.

\bibitem[{Yan and Yin(2012)}]{yan2012droidscope}
Yan, L.~K., Yin, H., 2012. Droidscope: seamlessly reconstructing the os and
  {Dalvik} semantic views for dynamic {Android} malware analysis. In: Presented
  as part of the 21st USENIX Security Symposium (USENIX Security 12). pp.
  569--584.

\bibitem[{Yang et~al.(2014)Yang, Xu, Gu, Yegneswaran, and
  Porras}]{yang2014droidminer}
Yang, C., Xu, Z., Gu, G., Yegneswaran, V., Porras, P., 2014. Droidminer:
  Automated mining and characterization of fine-grained malicious behaviors in
  {Android} applications. In: European Symposium on Research in Computer
  Security. Springer, pp. 163--182.

\bibitem[{Yang et~al.(2015)Yang, Xiao, Andow, Li, Xie, and
  Enck}]{yang2015appcontext}
Yang, W., Xiao, X., Andow, B., Li, S., Xie, T., Enck, W., 2015. Appcontext:
  Differentiating malicious and benign mobile {App} behaviors using context.
  In: Software Engineering (ICSE), 2015 IEEE/ACM 37th IEEE International
  Conference on. Vol.~1. IEEE, pp. 303--313.

\bibitem[{Zhang et~al.(2016)Zhang, Chan, Biggio, Yeung, and
  Roli}]{zhang2016adversarial}
Zhang, F., Chan, P.~P., Biggio, B., Yeung, D.~S., Roli, F., 2016. Adversarial
  feature selection against evasion attacks. IEEE transactions on cybernetics
  46~(3), 766--777.

\bibitem[{Zhang et~al.(2014)Zhang, Duan, Yin, and Zhao}]{zhang2014semantics}
Zhang, M., Duan, Y., Yin, H., Zhao, Z., 2014. Semantics-aware {Android} malware
  classification using weighted contextual {API} dependency graphs. In:
  Proceedings of the 2014 ACM SIGSAC Conference on Computer and Communications
  Security. ACM, pp. 1105--1116.

\bibitem[{Zhou et~al.(2013)Zhou, Zhou, Grace, Jiang, and Zou}]{zhou2013fast}
Zhou, W., Zhou, Y., Grace, M., Jiang, X., Zou, S., 2013. Fast, scalable
  detection of piggybacked mobile applications. In: Proceedings of the third
  ACM Conference on Data and Application Security and Privacy. ACM, pp.
  185--196.

\bibitem[{Zhou and Jiang(2012)}]{zhou2012dissecting}
Zhou, Y., Jiang, X., 2012. Dissecting android malware: Characterization and
  evolution. In: Security and Privacy (SP), 2012 IEEE Symposium on. IEEE, pp.
  95--109.

\bibitem[{Zhou et~al.(2012)Zhou, Wang, Zhou, and Jiang}]{zhou2012hey}
Zhou, Y., Wang, Z., Zhou, W., Jiang, X., 2012. Hey, you, get off of my market:
  detecting malicious apps in official and alternative android markets. In:
  NDSS. Vol.~25. pp. 50--52.

\end{thebibliography}

\newpage
\appendix
\section{Syntax and Semantic Features}\label{appendix-sec1}
\begin{table}[ht]\scriptsize \renewcommand \arraystretch{1.1}
\captionsetup{justification = centering}
\caption{175 syntax features and 20 semantic features for training classifiers.}
    \label{fig:features_whole}
 \begin{minipage}{\columnwidth}
 \begin{center}
 \begin{tabular}{|l|l|l|l|}
 \hline {\bf PERMISSION} 				& SET\_WALLPAPER & TelephonyManager.getSubscriberId &LocationManager.addNmeaListener \\
 \hline  ACCESS\_COARSE\_LOCATION 	 & SET\_WALLPAPER\_HINTS 	& TelephonyManager.getVoiceMailNumber &LocationManager.addProximityAlert \\
 \hline  ACCESS\_FINE\_LOCATION 		& STATUS\_BAR 	& TelephonyManager.hasIccCard  & LocationManager.addTestProvider\\
 \hline  {\tiny ACCESS\_LOCATION\_EXTRA\_COMMANDS} & SYSTEM\_ALERT\_WINDOW & TelephonyManager.isNetworkRoaming &LocationManager.clearTestProviderLocation\\
 \hline  ACCESS\_NETWORK\_STATE  		& UPDATE\_DEVICE\_STATS 	& SmsManager.divideMessage &LocationManager.getBestProvider\\
 \hline  ACCESS\_WIFI\_STATE 			& USE\_CREDENTIALS 	& SmsManager.getDefault & LocationManager.getGpsStatus\\
 \hline  AUTHENTICATE\_ACCOUNTS 	& VIBRATE 	& SmsManager.sendDataMessage &LocationManager.getLastKnownLocation\\
 \hline  BATTERY\_STATS 			         & WAKE\_LOCK 	& SmsManager.sendMultipartTextMessage &LocationManager.requestLocationUpdates\\
 \hline BLUETOOTH 					& WRITE\_APN\_SETTINGS & SmsManager.sendTextMessage &LocationManager.sendExtraCommand \\
 \hline BROADCAST\_SMS 				& WRITE\_SETTINGS & HttpURLConnection.disconnect &WifiManager.addNetwork  \\
 \hline BROADCAST\_STICKY 			& WRITE\_SMS & {\tiny HttpURLConnection.getContentEncoding} &WifiManager.calculateSignalLevel \\
 \hline CALL\_PHONE 				& {\tiny WRITE\_EXTERNAL\_STORAGE} & HttpURLConnection.getPermission &WifiManager.createWifiLock \\
 \hline CAMERA 						& {\bf INTENT} & HttpURLConnection.getRequestMethod &WifiManager.disconnect \\
 \hline {\tiny CHANGE\_COMPONENT\_ENABLED\_STATE}& action.DELETE & HttpURLConnection.getResponseCode &WifiManager.enableNetwork \\
 \hline CHANGE\_CONFIGURATION 		& action.GET\_CONTENT & {\tiny HttpURLConnection.getResponseMessage} &WifiManager.getConfiguredNetworks \\
 \hline CHANGE\_NETWORK\_STATE		& action.MAIN & HttpURLConnection.usingProxy&WifiManager.getConnectionInfo \\
 \hline {\tiny CHANGE\_WIFI\_MULTICAST\_STATE}& action.PICK & ContentResolver.bulkInsert&WifiManager.getDhcpInfo \\
 \hline CHANGE\_WIFI\_STATE			& action.SEND &ContentResolver.getType &WifiManager.getScanResults \\
 \hline CLEAR\_APP\_CACHE			& action.SET\_WALLPAPER &ContentResolver.openAssetFileDescriptor &WifiManager.getWifiState \\
 \hline {\tiny CONTROL\_LOCATION\_UPDATES}	& action.VIEW &ContentResolver.query &WifiManager.isWifiEnabled \\
 \hline DELETE\_CACHE\_FILES			& category.BROWSABLE &ContentResolver.registerContentObserver &WifiManager.removeNetwork \\
 \hline DELETE\_PACKAGES				& category.DEFAULT &ContentResolver.update & WifiManager.saveConfiguration\\
 \hline DEVICE\_POWER				& category.HOME &ContentResolver.delete &WifiManager.setWifiEnabled \\
 \hline DISABLE\_KEYGUARD			& category.INFO &Runtime.getRuntime &NotificationManager.cancel \\
 \hline EXPAND\_STATUS\_BAR			& category.LAUNCHER &Runtime.exec &NotificationManager.notify \\
 \hline FLASHLIGHT					& {\bf HARDWARE} &Runtime.addShutdownHook &PackageManager.checkPermission \\
 \hline GET\_PACKAGE\_SIZE			& camera &Runtime.maxMemory &PowerManager.isInteractive \\
 \hline GET\_TASKS					& camera.autofocus &URLConnection.addRequestProperty &PowerManager.isScreenOn \\
 \hline INSTALL\_PACKAGES			& sensor.accelerometer &URLConnection.connect &PowerManager.newWakeLock \\
 \hline INTERNET					& telephony &URLConnection.getConnectTimeout &{\bf SEMANTIC} \\
 \hline {\tiny KILL\_BACKGROUND\_PROCESSES}& touchscreen &URLConnection.getContent & *Install application* \\
 \hline MODIFY\_PHONE\_STATE		& {\bf API CALL} &URLConnection.getContentType &*Uninstall application* \\
 \hline {\tiny MOUNT\_UNMOUNT\_FILESYSTEMS}& URL.openConnection &URLConnection.getDefaultUseCaches & *Get installed packages*\\
 \hline NFC						& URL.openStream &URLConnection.getPermission &*Monitor URI* \\
 \hline PERSISTENT\_ACTIVITY			& URL.getContent &URLConnection.getURL &*Download file* \\
 \hline PROCESS\_OUTGOING\_CALLS	& TelephonyManager.getCallState  &URLConnection.setConnectTimeout &*Get location* \\
 \hline READ\_CALL\_LOG				& {\tiny TelephonyManager.getCellLocation} &URLConnection.setReadTimeout & *Read SD card*\\
 \hline READ\_CONTACTS				&TelephonyManager.getDeviceId  & ActivityManager.getLargeMemoryClass&*Write SD card* \\
 \hline READ\_EXTERNAL\_STORAGE		&{\tiny TelephonyManager.getDeviceSoftwareVersion}  &{\tiny ActivityManager.getRunningAppProcesses}&*Request for chmod* \\
 \hline READ\_LOGS					&{\tiny TelephonyManager.getNeighboringCellInfo}  & ActivityManager.isLowRamDevice&*Start http connection* \\
 \hline READ\_PHONE\_STATE			&{\tiny TelephonyManager.getNetworkCountryIso}  &{\tiny ActivityManager.killBackgroundProcesses} &*Send Sms* \\
 \hline READ\_PROFILE				&{\tiny TelephonyManager.getNetworkOperator}  &ActivityManager.restartPackage &*Receive Sms* \\
 \hline READ\_SMS					&{\tiny TelephonyManager.getNetworkOperatorName}  &BluetoothAdapter.cancelDiscovery &*Delete Sms* \\
 \hline RECEIVE\_BOOT\_COMPLETED	&{\tiny TelephonyManager.getNetworkType} &BluetoothAdapter.getAddress &*Intercept Sms receiver* \\
 \hline RECEIVE\_MMS					&{\tiny TelephonyManager.getPhoneType}  &BluetoothAdapter.getBondedDevices &*Get wifi info* \\
 \hline RECEIVE\_SMS					&{\tiny TelephonyManager.getSimCountryIso}  &BluetoothAdapter.getRemoteDevice &*Get Logs* \\
 \hline RECEIVE\_WAP\_PUSH			&{\tiny TelephonyManager.getSimOperator} &BluetoothSocket.connect &*Get Class loader* \\
 \hline RECORD\_AUDIO				&{\tiny TelephonyManager.getSimOperatorName}  &DownloadManager.enqueue & *Get contacts*\\
 \hline RESTART\_PACKAGES			&{\tiny TelephonyManager.getSimSerialNumber}  &DownloadManager.query & *Get account*\\
 \hline SEND\_SMS 					&TelephonyManager.getSimState  &LocationManager.addGpsStatusListener &{\tiny *Get phone type/Sim serial number/device id/subscriber id/IMSI*} \\
 \hline
\end{tabular}
\end{center}
\end{minipage}
\end{table}

\clearpage
\section*{\large Biography}
\noindent \textbf{Sen Chen} is pursuing his Ph.D. degree at the School of Computer Science and Software Engineering of East China Normal University, focusing primarily on areas of smartphone security, Android malware, vulnerability and program analysis. 
He has received the MobiCom 2016 Travel Grant Award. He is currently serving as a visiting scholar in Cyber Security Lab at Nanyang Technological University. He is currently advised by Professor Lihua Xu (ECNU) and Yang Liu (NTU).
\vspace{2mm}
  
\noindent \textbf{Minhui Xue} is pursuing his Ph.D. degree at the School of Computer Science and Software Engineering of East China Normal University. He is also serving as a visiting scholar at the Courant Institute of Mathematical Sciences and Tandon School of Engineering at New York University, as well as a research assistant at New York University Shanghai, advised by Professor Keith W. Ross (NYU and NYU Shanghai). Previously, he received a Bachelor of Science degree in the field of fundamental mathematics from East China Normal University in July 2013. His current research interests are in data-driven analysis of online social networks and privacy, focusing primarily on computer science and mathematics. 
\vspace{2mm}

\noindent \textbf{Lingling Fan} is pursuing her Ph.D. degree at the School of Computer Science and Software Engineering of East China Normal University, focusing on software testing, model checking, and Android application analysis. She is interested in software testing and analysis, aiming at bug revelation and bug localization, and malware detection. She received a Bachelor of Science degree in the field of computer science and technology from ECNU, as an excellent graduate student of Shanghai. She received an Excellent Student award from ECNU. She is currently advised by Professor Lihua Xu (ECNU) and Yang Liu (NTU).
\vspace{2mm}

\noindent \textbf{Shuang Hao} is an Assistant Professor in the Department of Computer Science at the University of Texas at Dallas. He is broadly interested in all aspects of network and system security. His work follows a measurement and data driven approach to characterize and detect critical security issues in large-scale systems. His current research focuses on anomaly detection, underground economics, DNS analysis, web and mobile security. He obtained his Ph.D. in Computer Science at the Georgia Institute of Technology, and he did a Postdoc at the University of California, Santa Barbara. 
\vspace{2mm}

\noindent \textbf{Lihua Xu} is an Associate Professor with School of Computer Science and Software Engineering, East China Normal University. She received her Ph.D. and M.Sc. degree from University of California at Irvine. Her current research focuses on software engineering, automated software analysis and testing, and mobile security. 
\vspace{2mm}

\noindent \textbf{Haojin Zhu} is currently a Professor with Department of Computer Science and Engineering, Shanghai Jiao Tong University, China. He received his B.Sc. degree (2002) from Wuhan University (China), his M.Sc.(2005) degree from Shanghai Jiao Tong University (China), both in computer science and the Ph.D. in Electrical and Computer Engineering from the University of Waterloo (Canada), in 2009. His current research interests include network security and data privacy. He serves as the Associate/Guest Editor of IEEE Internet of Things Journal, IEEE Wireless Communications, IEEE Network, and Peer-to-Peer Networking and Applications.
\vspace{2mm}

\noindent \textbf{Bo Li} is currently a Postdoctoral Fellow at the EECS department at UC Berkeley University, working with Prof. Dawn Song. She will join the EECS department at University of Illinois at Urbana -- Champaign as an Assistant Professor in June 2018. Her research focuses on both theoretical and practical aspects of machine learning, security, privacy, game theory, social networks, and adversarial deep learning.

\end{document}